\documentclass{article}

\usepackage{arxiv}

\usepackage[utf8]{inputenc} % allow utf-8 input
\usepackage[T1]{fontenc}    % use 8-bit T1 fonts    % hyperlinks
\usepackage{url}            % simple URL typesetting
\usepackage{booktabs}       % professional-quality tables
\usepackage{amsfonts}       % blackboard math symbols
\usepackage{nicefrac}       % compact symbols for 1/2, etc.
\usepackage{microtype}      % microtypography
\usepackage{lipsum}		% Can be removed after putting your text content
\usepackage{graphicx}
\usepackage{natbib}
\usepackage{doi}
\usepackage{newtxtext}
\usepackage{newtxmath}
\usepackage{authblk}
\usepackage{amsmath} 
\usepackage{subeqnarray}

\title{Wake dynamics of finite-aspect-ratio rotating circular cylinders at low Reynolds number}

\author[1]{Kai Zhang}
\author[1]{Yong Cao}
\author[2]{Hanfeng Wang}
\author[1]{Yan Bao}
\author[1]{Bin Zhao}
\author[1,3]{Dai Zhou}
\affil[1]{State Key Laboratory of Ocean Engineering, School of Ocean and Civil Engineering, Shanghai Jiao Tong University, Shanghai 200240, China}
\affil[2]{School of Civil Engineering, Central South University, Changsha 410083, China}
\affil[3]{Shenzhen Research Institute of Shanghai Jiao Tong University, Shenzhen 518063, China}

\begin{document}
\maketitle

\begin{abstract}
We perform direct numerical simulations of flows over finite-aspect-ratio rotating circular cylinders at a Reynolds number of 150 over a range of aspect ratios ($AR=2-12$) and rotation rates ($\alpha=0-5$), aiming at revealing the free-end effects on the wake dynamics and aerodynamic performance. 
As a direct consequence of lift generation, a pair of counter-rotating tip vortices is formed at the free ends.
At low rotation rates, the finite rotating cylinder behaves like a typical bluff body that generates unsteady vortex shedding with three-dimensional modal structures.
Such unsteady flows can be stabilized not only by increasing rotation rate that weakens the free shear layer, but also by decreasing aspect ratio which enhances the tip-vortex-induced downwash.
A further increase of $\alpha$ triggers the onset of unsteadiness in the tip vortices.
At still higher rotation rates, the C-shaped Taylor-like vortices bounded on the cylinder surface emerge from the free ends and propel towards the midspan due to the self-induced velocity by vortex-wall interaction.
With increasing $\alpha$, the free-end effects penetrate to the inboard span, leading to reduced lift and elevated drag compared to the two-dimensional flows.
The three-dimensional end effects can be effectively suppressed by the addition of end plates, which position the tip vortices away from the cylinder, thereby significantly improving the aerodynamic performance.
This study reveals the mechanisms for the formation of three-dimensional wakes under the influence of the free ends of finite rotating cylinders. 
The insights obtained here can serve as a stepping stone for understanding the complex high-$Re$ flows that are more relevant to industrial applications.\end{abstract}

% keywords can be removed
%\keywords{First keyword \and Second keyword \and More}

\section{Introduction}
\label{sec:introduction}
The flow over rotating circular cylinders has been a celebrated topic of research due to its fundamental connection to the Magnus effect \citep{Magnus1853,Borg1986Magnus,anderson2011fundamentals}. 
This canonical flow configuration serves as a critical testbed for diverse scientific and engineering pursuits, including wind-assisted propulsion \citep{flettner1925,seifert2012review,sedaghat2014magnus} and advanced flow control techniques \citep{schulmeister2017flow,fan2020reinforcement,yu2020turbulent,bao2022laminar,wu2022transition}.
The enduring interest in this flow is further amplified by the remarkably rich wake dynamics it exhibits.

The wake dynamics of the flow around rotating cylinders can be complex even in the two-dimensional setting \citep{Stojkovic2002pof,Stojkovic2003pof,mittal2003flow,PADRINO_JOSEPH_2006,pralits2010instability}.
At low rotation rates, the cylinder behaves like a bluff body, resulting in shedding mode I characterized by K\'arm\'an vortex street.
As the rotation rate increases, the shear layer on the retreating side of the rotating cylinder weakens.
This leads to the quenching of shedding mode I and the rise of the steady flow regime, which has a tail-like structure in the wake.
However, beyond a certain threshold, the two-dimensional flow exhibits a re-emergence of unsteadiness in a small window of higher rotation rates.
This unsteady periodic flow, also termed shedding mode II, is characterized by single-sided shedding with a frequency much lower than that of shedding mode I.
%However, the unsteady shedding mode II cannot be observed in laboratory experiments due to the inherent three-dimensionality of this flow.
Upon further increasing the rotation rate, the wake again transitions to a steady flow state, in which the shear layer wraps around the cylinder.
Interestingly, there exist multiple solutions for the steady flow state at high rotation rates \citep{pralits2010instability,Rao2013jfm_a,Thompson_et_al_2014}.
A comprehensive analysis of the complex bifurcation scenario for the two-dimensional laminar flow past rotating cylinders is provided by \citet{sierra2020bifurcation} and \citet{brons2021organizing}.

Both steady and unsteady two-dimensional flows over rotating cylinders are susceptible to the intrinsic three-dimensional instabilities as the Reynolds number and rotation rate vary \citep{Radi2013JFM,Rao2013jfm_b,Rao2013jfm_a,pralits2013three,rao2015review,navrose2015three}. 
For unsteady shedding mode I, mode A and mode B instabilities can develop on top of the K\'arm\'an vortex shedding at low rotation rates, similar to the wake of a non-rotating cylinder as described by \citet{williamson1996ARFM}.
At higher rotation rates, a subharmonic instability termed mode C appears as the wake becomes clearly asymmetric.
The associated vortical structures are shifted by half a wavelength along the spanwise direction over each base flow period, indicating the subharmonic nature of this mode.
Additionally, mode D appears within the strained vortices of the wake, occupying a narrow region between mode C and the transition to steady flow.

For the steady base flows at high rotation rates, three-dimensionality manifests through mode E and mode F instabilities. 
Mode E appears as steady streamwise vortices with a characteristic spanwise wavelength of approximately $2D$, similar to Mode D. 
The physical mechanism for Mode E is hypothesized to be an amplification of perturbations in the high-strain region near the hyperbolic point at the rear of the single recirculation zone.
Mode F has a much shorter characteristic spanwise wavelength of approximately $0.45D$, and occurs at an even higher rotation rate than Mode E.
The vortex filaments grow primarily in the boundary layer of the spinning cylinder, and wrap partly around the cylinder and extend into the wake.
Linear stability analysis indicates that mode F can be understood as a centrifugal instability \citep{Taylor1923PTRS,Drazin_Reid_2004} of the closed region of flow near the cylinder surface.
A notable characteristic of mode F is that its vortical structures move along the cylinder span, behaving like a traveling wave \citep{Radi2013JFM}.
The diversity of these intrinsic three-dimensional instability modes, driven by different physical mechanisms, reveals the rich complexity underlying rotating cylinder wake dynamics.

In engineering applications like fuel-saving rotor sails on vessels, the wake dynamics of rotating cylinders are inevitably influenced by end boundary conditions. These boundaries act as extrinsic sources that drive the development of three-dimensional flows. The effects of these end boundary conditions on wake behavior have been extensively investigated in non-rotating cylindrical structures \citep{slaouti1981experimental,ramberg1983effects,williamson1989oblique,park2000free,roh2003vortical,inoue2008vortex,WANG_ZHOU_2009,KRAJNOVIC_2011,sumner2013flow,mittal2021cellular,cao2022jfm,zhang2023OE}.
The free end of the cylinder generally induces the downward flow from the free end to the central wake.
The ground effects lead to the formation of horseshoe vortices at the junctions of the cylinder and ground, and can induce upward flow similar to the free end.
Under these effects, the vortex shedding process exhibits a range of intricate flow phenomena including oblique shedding, vortex dislocation, cellular shedding and frequency modulation.

The end effect is also an important consideration for lifting surfaces like wings.
For airfoils, finite wing theory has long been established to account for tip vortex effects on the aerodynamic performance of high-speed aircraft \citep{anderson2011fundamentals}. 
Recent interest in low-altitude micro air vehicles has also generated considerable research into separated flows over finite wings \citep{taira2009three,DeVoria_Mohseni_2017,zhang2020laminar,zhang2020formation,zhang2022laminar,pandi2023JFM,smith2024effect,Zhu_Wang_Liu_2024}.
These studies revealed that the downwash effects induced by the tip vortices can stabilize leading-edge vortices on low-aspect-ratio wings, thereby providing enhanced vortex lift and improved aerodynamic performance.
The literature on both bluff bodies and lifting surfaces consistently demonstrate that three-dimensional flows under the influence of end boundary conditions exhibit markedly different characteristics compared to their two-dimensional counterparts.

The focus of the present study is the three-dimensional wake dynamics of finite-aspect-ratio rotating circular cylinders, which bear resemblance to both non-rotating bluff bodies and wings, but have received much less attention in the literature. 
Among the few who have addressed this problem, \citet{mittal2004JAM} studied the flows over rotating cylinders confined by no-slip walls. It was revealed that although the two-dimensional flow at the same configuration is stable, the inhomogeneous end boundary induces the centrifugal instabilities along the span. In addition, the presence of a no-slip side wall can result in flow separation near the cylinder ends, leading to a loss in lift and an increase in drag.
\citet{yang2023JFM} covered the wake dynamics of short, rotating cylinders using direct numerical simulation (DNS) and global stability analysis. They showed that rotation has a non-monotonic effect on flow stability: moderate rotation stabilizes the flow by suppressing vortex shedding, whereas strong rotation destabilizes it by introducing new, complex, low-frequency instabilities. 
\citet{massaro2024direct} carried out DNS of the flows over a wall-mounted cylinder at $Re=3000$, revealing the large-scale vortical motion and its profound impact on drag, and highlighting the critical importance of considering the cylinder-wall interaction for accurate performance prediction and design optimization in practical applications.
On the experimental side, a series of wind tunnel tests have been carried out to study the aerodynamic performance of rotor sails, mainly focusing on the aerodynamic forces \citep{bordogna2019experiments,ma2022jweia} and multi-rotor wake interactions \citep{BORDOGNA2020JWEIA,chen2023experimental}.

Despite the extensive characterization of end-effect phenomena in stationary cylinders and finite wings, these mechanisms remain inadequately resolved for finite rotating cylinders.
Specifically, the fundamental physical processes through which free-end conditions dictate the formation, organization, and evolution of three-dimensional vortical structures are poorly understood. Consequently, their impact on aerodynamic forces lacks quantitative assessment. 
To address these questions, we perform a series of direct numerical simulations of flows over finite rotating cylinders at a low Reynolds number.
By systematically varying the rotation rate and aspect ratio, we aim to identify the dominant flow regimes and transition boundaries, elucidate the mechanisms governing wake vortical structure evolution under different parameters, quantify the resulting impact on lift and drag characteristics, and establish a comprehensive understanding of three-dimensional wake dynamics in rotating finite cylinders.
In what follows, we present the computational setup and its validation in \S \ref{sec:setup}.
The results are discussed in \S \ref{sec:results}, including the formation of tip vortices, classification and characterization of different wake types, analysis of aerodynamic forces, and assessment of end plate effects on wake behavior. 
Finally, we summarize our key findings in \S \ref{sec:conclusions}.

\section{Computational setup}
\label{sec:setup}
\subsection{Problem description}

\begin{figure}
\centering
\includegraphics[width=0.9\textwidth]{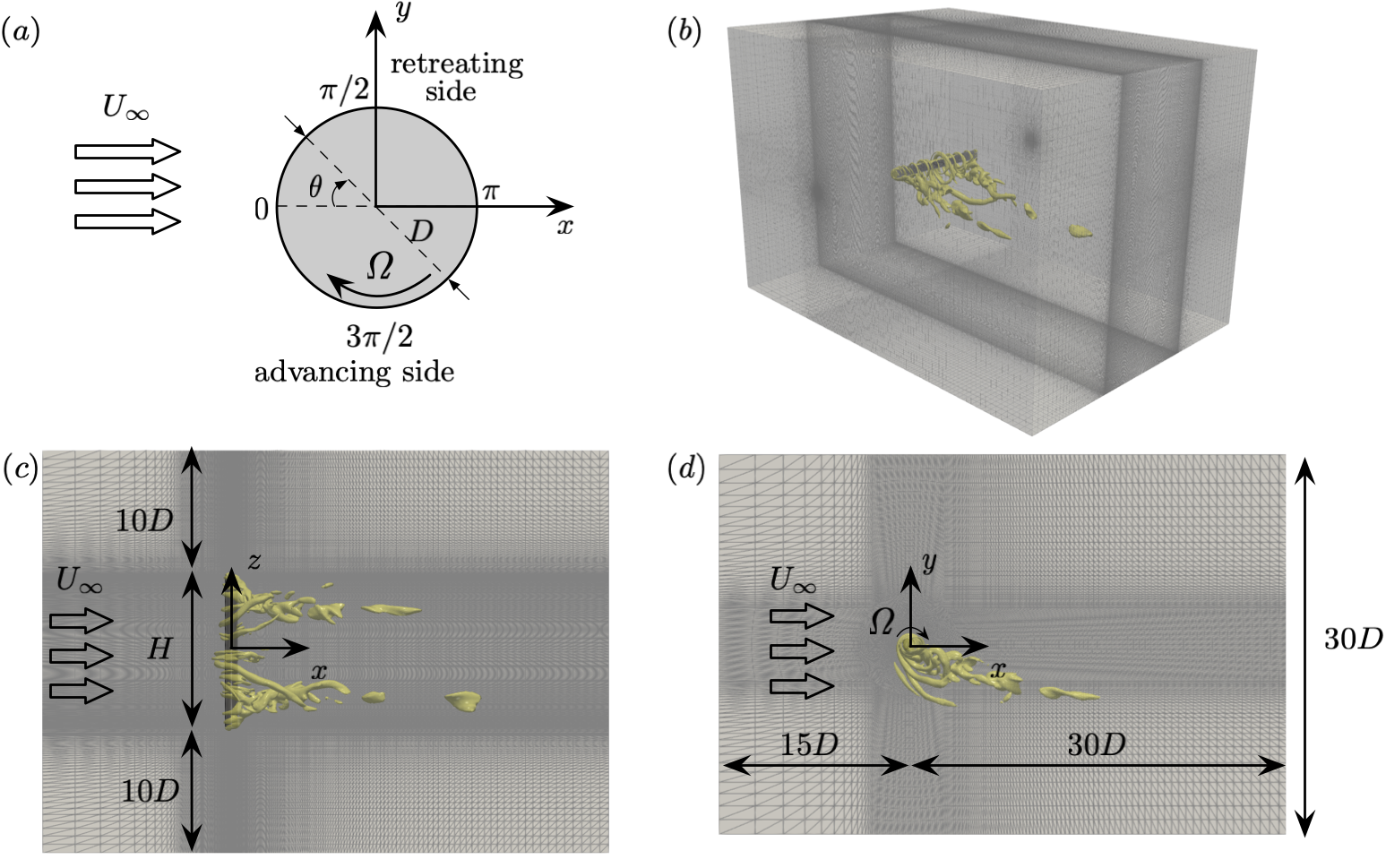}
\caption{The computational setup and mesh design for flow over a rotating finite cylinder. ($a$) problem setup, ($b$--$d$) a perspective view, top view, and side view of the computational domain and mesh. Shown is the case of $(AR,\alpha)=(12,5)$. The vortical structures are visualized by isosurfaces of $Q=1$ in yellow, where $Q$ is the second invariant of the velocity gradient tensor.}
\label{fig:Mesh}
\end{figure}

We carry out direct numerical simulations of incompressible flows over rotating cylinders with finite lengths, as shown in figure \ref{fig:Mesh}.
The circular cylinder with a diameter $D$ and spanwise length $H$ is subjected to a uniform inflow with velocity $U_{\infty}$ in the $x$ direction. 
The cylinder rotates in the clockwise direction along its axial axis $z$ so that it generates a positive lift force in the $y$ direction. 
The aspect ratio of the finite cylinder is defined as $AR=H/D$, and is varied from 2 to 12 in this study.
Denoting the angular velocity of the rotation as $\Omega$, the nondimensional rotation rate of the cylinder is expressed as $\alpha=\Omega R/U_{\infty}$ (where $R=D/2$ is the radius of the cylinder), and ranges from 0 to 5. 
The side where the cylinder's surface velocity is parallel to the freestream is termed the retreating side ($\theta = \pi/2$). The opposite side, where the surface velocity opposes the freestream, is termed the advancing side ($\theta = 3\pi/2$).

The Reynolds number, $Re = U_{\infty}D/\nu$ (where $\nu$ is the kinematic viscosity of fluid), is fixed at 150, ensuring laminar flow while allowing various flow patterns to develop.
Hereafter, we report the variables in their nondimensional forms by normalizing spatial quantities with $D$, velocity with $U_{\infty}$, time with $D/U_{\infty}$, and frequency with $U_\infty/D$.

\subsection{Numerical methods}
\label{sec:methods}
The three-dimensional flows over the rotating finite cylinder are studied by numerically solving the incompressible Navier--Stokes (N-S) equations:
\begin{subeqnarray}
 \frac{\partial \boldsymbol{u}}{\partial t} +  \boldsymbol{u} \cdot \boldsymbol{\nabla} \boldsymbol{u} & = & -\boldsymbol{\nabla} p +\displaystyle{\frac{1}{Re}} \boldsymbol{\nabla}^2 \boldsymbol{u},\\
 \boldsymbol{\nabla} \cdot \boldsymbol{u} & = & 0,
\label{equ:NS}
\end{subeqnarray}
where $\boldsymbol{u}=(u_x,u_y,u_z)$ is the velocity vector and $p$ is the pressure.
The finite-volume-based flow solver pimpleFoam from the open-source CFD toolbox OpenFOAM is used for solving the three-dimensional N--S equations with second-order accuracy in both space and time. 
The rectangular computational domain covers $(x,y,z)\in [-15D, 30D]\times [-15D, 15D]\times [-(H/2+10D), (H/2+10D)]$, as shown in figure \ref{fig:Mesh}. 
This results in a maximum blockage ratio of 1.25\% for the cases with $AR=12$. 
The inlet is prescribed with a uniform velocity in the $x$ direction $U_\infty$.
The surfaces of the cylinders are treated as no-slip so that the boundary layer flow move along with the linear velocity of the cylinder surface. 
The side boundaries are considered as slip. 
The zero-gradient condition is applied to the outlet for velocity, where a reference pressure $p=0$ is specified. 
The maximum number of control volumes amounts to around $1.6\times 10^7$ for cases with $AR=12$. The mesh is clustered around and in the near wake of the cylinder, as shown in figure \ref{fig:Mesh}($b$--$d$).

The lift and drag forces are reported in their nondimensional forms:
\begin{subeqnarray}
C_L &= &\displaystyle{\frac{\displaystyle{\int}_S(-p\boldsymbol{n}+\boldsymbol{\tau})\cdot \boldsymbol{e_y}\mathrm{d}S}{U_{\infty}^2HD/2}},\\
C_D &= &\displaystyle{\frac{\displaystyle{\int}_S(-p\boldsymbol{n}+\boldsymbol{\tau})\cdot \boldsymbol{e_x}\mathrm{d}S}{U_{\infty}^2HD/2}},
\label{equ:forceCoeffs}
\end{subeqnarray}
where $S$ denotes the cylinder surface, $\boldsymbol{n}$ is the unit normal vector pointing outward from the cylinder surface. 
Unit vectors $\boldsymbol{e_y}$ and $\boldsymbol{e_x}$ point in the lift and drag directions, respectively, and $\boldsymbol{\tau}=\mu\boldsymbol{\omega}\times \boldsymbol{n}$ is the skin-friction vector ($\mu$ is the dynamic viscosity of fluid and $\boldsymbol{\omega}=\boldsymbol{\nabla}\times \boldsymbol{u}$ is the vorticity vector).

\subsection{Validation}
\label{sec:validation}
\begin{figure}
\centering
\includegraphics[width=0.95\textwidth]{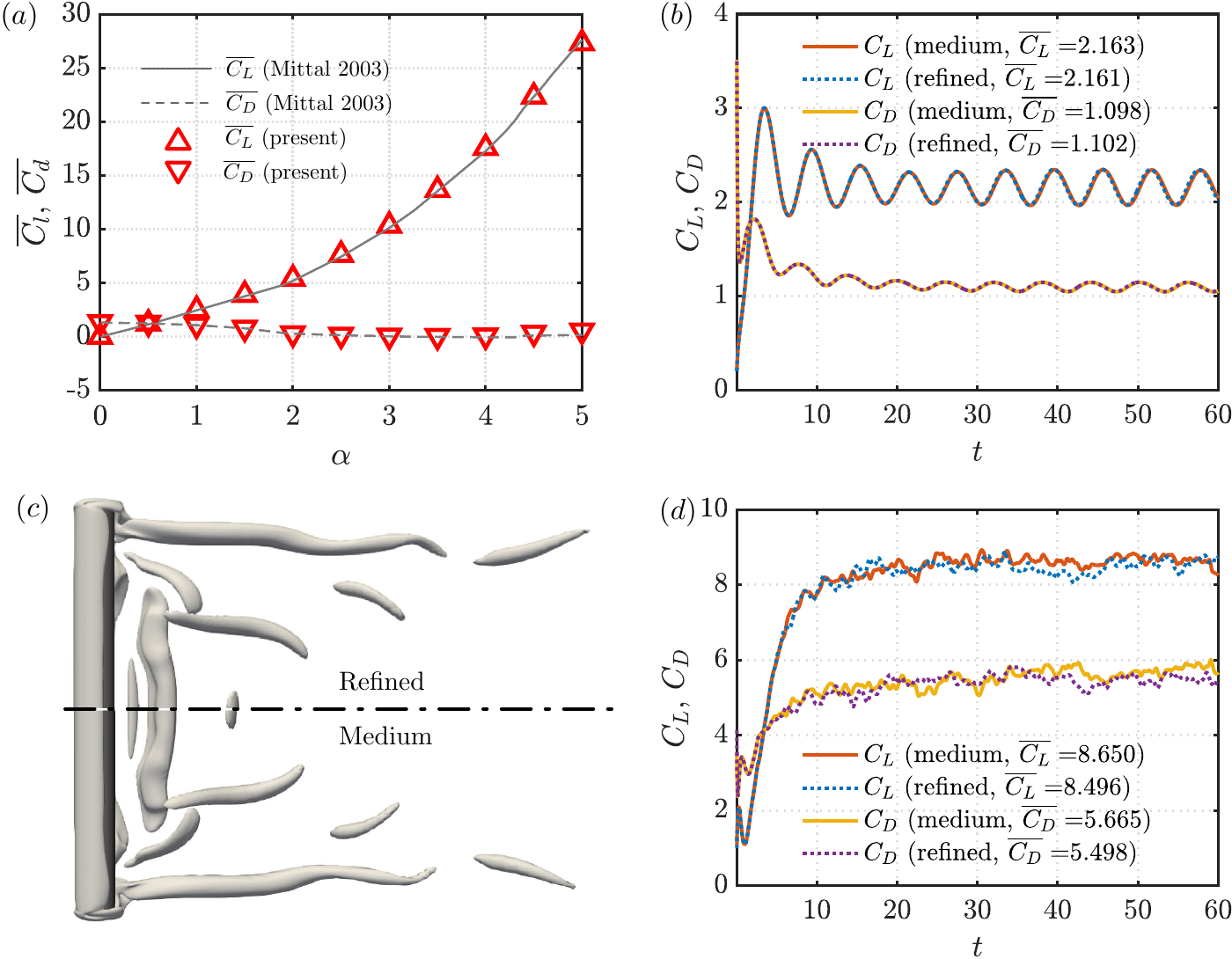}
\caption{Validation of the numerical setup. ($a$) Comparison of the mean lift and drag force coefficients from 2-D simulations between our results and those from \citet{mittal2003flow}, ($b$) comparison of the time histories of aerodynamic forces for $(AR,\alpha)=(10,1)$ between medium and refined mesh, ($c$) comparison of the vortical structures visualized by isosurfaces of $Q=1$ for $(AR,\alpha)=(10,1)$ and $(d)$ comparison of the time histories of aerodynamic forces for $(AR,\alpha)=(10,5)$.}
\label{fig:Validation}
\end{figure}
We first validate the numerical setup by simulating two-dimensional flow over a rotating cylinder at $Re=200$. The results are compared with the benchmark study by \citet{mittal2003flow}. As shown in figure \ref{fig:Validation}($a$), both studies predict a monotonic increase of lift with increasing rotation rate $\alpha$, as well as the nearly zero drag.
We further conduct a mesh dependency test for the three-dimensional flows over finite rotating cylinders with $AR=10$ at $Re=150$. 
As shown in table \ref{tab:MeshTest}, two sets of meshes are designed, with the grid resolution considerably enhanced in the circumferential, spanwise, and radial directions of the cylinder in the refined mesh. The time-step is also reduced in the simulations with the refined mesh.
We select two rotation rates, $\alpha=1$ and 5, for the mesh dependency test. 
In the first case ($\alpha=1$), the temporal traces of the lift and drag coefficients match closely between the medium and refined mesh, and the wake vortical structures show no discernible difference as presented in figure \ref{fig:Validation}($b,c$).
In the second case with the largest rotation rate $\alpha=5$, the aerodynamic force coefficients from the two meshes overlap in the beginning phase of the simulation, but diverge after $t\approx 5$ due to the chaotic nature of the wake flow. 
Nevertheless, it can be appreciated that the two traces of force coefficients are close to each other in the time-averaged sense. 
From the above discussions, it is concluded that the medium grid resolution is sufficient for the current study, thus is used to carry out the simulations as reported next.

\begin{table}  
\setlength{\tabcolsep}{10pt}
 \begin{center}
  \begin{tabular}{lccccc}
    Mesh  & $N_{\theta}$   &   $N_z$ & $\Delta r_1$ & $N_{\textrm{CV}}$ & $\Delta t$   \\[3pt]
       medium   & 180 & 300 & 0.0092 & $1.38\times 10^7$ & 0.0025\\
       refined  & 240 & 400 & 0.0057 & $2.45\times 10^7$ & 0.002\\
  \end{tabular}
  \caption{
  Setups of the different meshes for the case of $(\alpha,AR)=(5, 10)$. $N_{\theta}$ and $N_z$ are the numbers of grid points on the cylinder surface and along the spanwise direction. $\Delta r_1$ is the height of the first layer of mesh on the cylinder surface. $N_{CV}$ is the total number of control volumes, and $\Delta t$ is the time step.}
\label{tab:MeshTest}
 \end{center}
\end{table}

\section{Results}
\label{sec:results}
We begin the discussions of results by examining the characteristics of tip vortices influenced by the rotation rate in \S \ref{sec:tipVortex}.
Subsequently, we present the different types of wake vortical structures that develop under the influence of these tip vortices in \S \ref{sec:overview}. 
The aerodynamic forces of the finite rotating cylinders are then analyzed in \S \ref{sec:forces}. 
Finally, we demonstrate the effectiveness of the end plates in mitigating the three-dimensional effects in finite rotating cylinder wakes in \S \ref{sec:endplates}.

\subsection{Tip vortices}
\label{sec:tipVortex}

\begin{figure}
\centering
\includegraphics[width=0.8\textwidth]{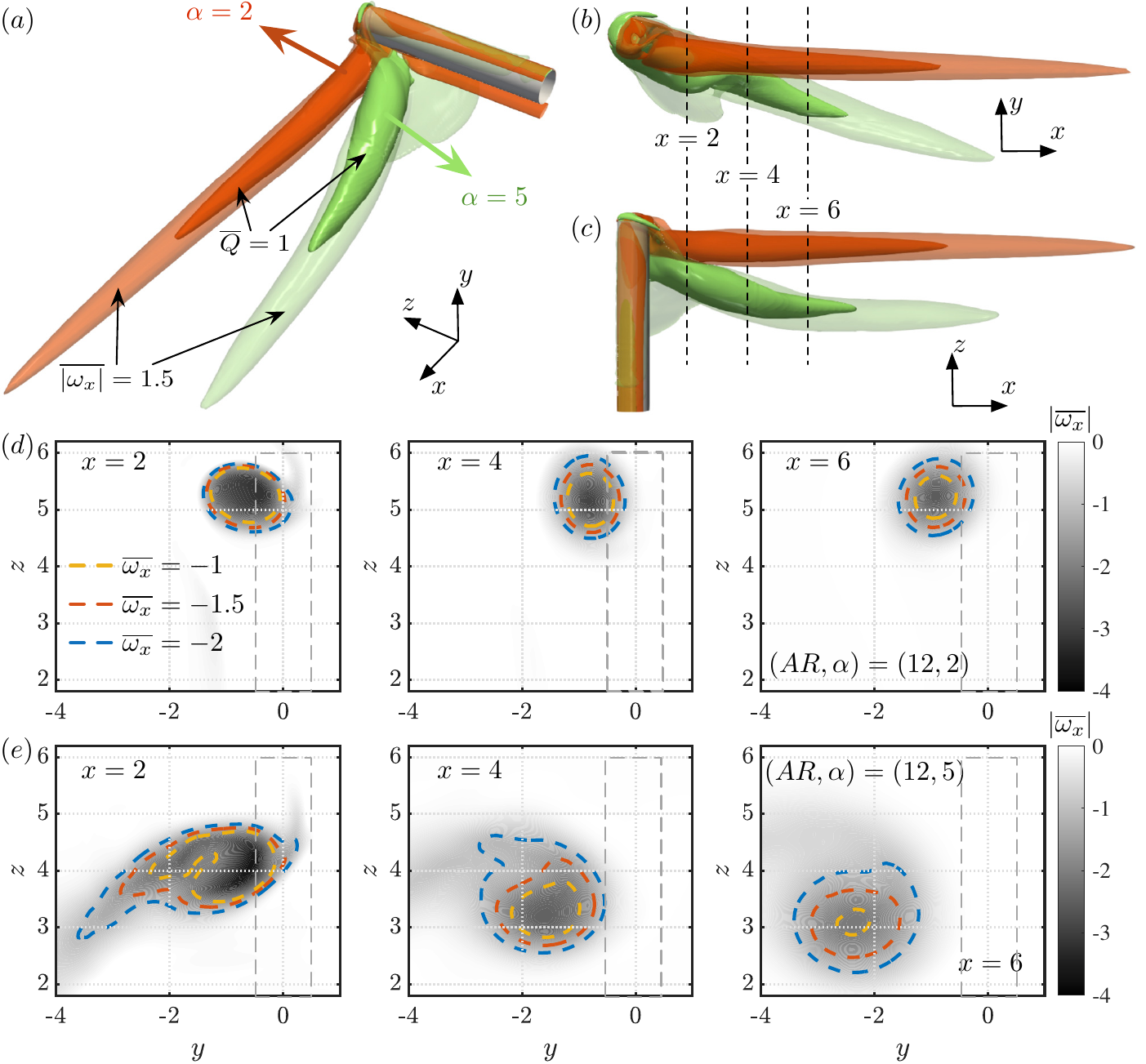}
\caption{Comparison of the time-averaged tip vortical structures between $\alpha=2$ (orange) and $\alpha=5$ (green) for a fixed $AR=12$. ($a$ -- $c$) Perspective view, side view, and top view of the tip vortices visualized by isosurfaces of $\overline{Q}=1$ (solid) and $\overline{|\omega_x|}=1.5$ (transparent), respectively. ($d$ -- $e$) Time-averaged streamwise vorticity fields at $x=2$, 4 and 6 for $\alpha=2$ and $\alpha=5$. The gray dashed boxes in $(d)$ and $(e)$ represent the location of the cylinders.}
\label{fig:TipVortex}
\end{figure}

As a classical lifting body, a finite rotating cylinder generates tip vortices due to the pressure difference between the advancing and retreating sides at the free end—a phenomenon analogous to that observed in finite wings \citep{schlichting2013aerodynamik,anderson2011fundamentals}.
To investigate the influences of rotation rate on the characteristics of the tip vortices, we choose the cases with the largest aspect ratio $AR=12$, so that the interactions between the counter-rotating tip vortices at each end of the cylinder are minimized.
We compare the tip vortices between two representative cases, $\alpha=2$ at which the wake is steady and $\alpha=5$ where the tip vortex breaks down. 
The time-averaged tip vortices are visualized using the isosurfaces of $\overline{Q}=1$ and $|\overline{\omega_x}|=1.5$ in figure \ref{fig:TipVortex}($a-c$). 
The contours of $\overline{\omega_x}=-1, -1.5$ and -2 are also used to delineate the shape of the vortex cores for the two cases in figure \ref{fig:TipVortex}($d,e$).

For the steady flow at $\alpha=2$, the tip vortex is featured by a straight and elongated vortical structured formed in the close vicinity of the free end.
Cross-sectional analysis on the $y$-$z$ planes reveals an initially elliptical shape in the near wake that gradually evolves into a more circular configuration as the vortex convects downstream.
In contrast, the time-averaged tip vortex for $\alpha=5$ is considerably shorter due to vortex breakdown (to be discussed in the next subsection).
For this case, the cross-sectional shape in the near wake exhibits greater irregularity, indicating that multiple vorticity sources likely contribute to tip vortex formation. 
Further downstream, while the vortex cross-section becomes more regular in shape, it remains significantly more diffused as a consequence of the breakdown process.
A notable distinction between the two cases is that the tip vortex at $\alpha=5$ exhibits greater inboard tilting compared to $\alpha=2$, suggesting a more pronounced influence on the inboard flow field.

\begin{figure}
\centering
\includegraphics[width=0.5\textwidth]{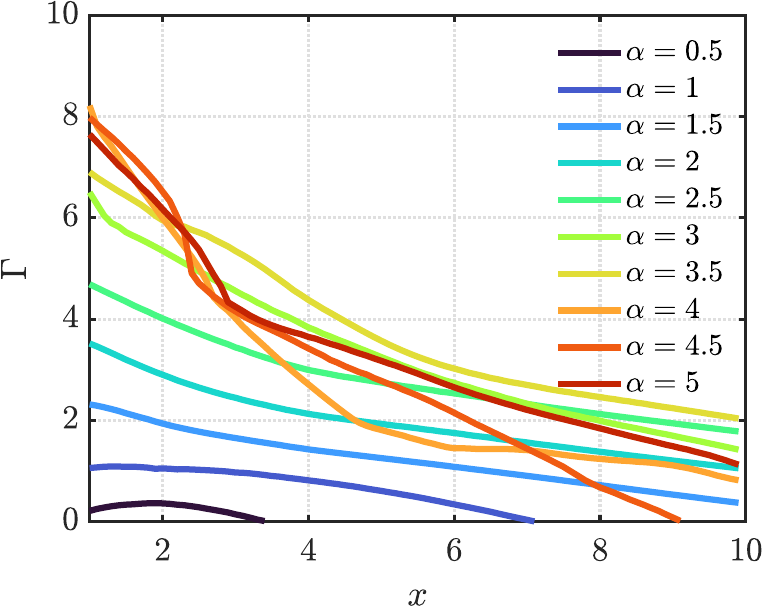}
\caption{Tip vortex circulation calculated based on contours of $|\overline{\omega_x}|=1.5$ for cases with $AR=12$.}
\label{fig:TipCirculation}
\end{figure}

We further quantify the streamwise evolution of tip vortex strength by evaluating the circulation $\Gamma=\oint_{C} \boldsymbol{u}\cdot \mathrm{d}\boldsymbol{l}$ on $y$-$z$ planes, where the integration contour $C$ is defined as the contour line of $\overline{\omega_x}=-1.5$. 
As shown in figure \ref{fig:TipCirculation}, circulation at low rotation rates decays nearly linearly along the streamwise direction. 
With increasing rotation rate, tip vortices intensify and extend farther downstream. 
This intensification plateaus for $\alpha\gtrsim 4$, where vortex breakdown occurs. 
Beyond this threshold, circulation drops drastically in the near wake before transitioning to a more gradual decay in the far wake. 
Consequently, high-$\alpha$ cases can exhibit shorter tip vortex lengths than their low-$\alpha$ counterparts, as observed in figure \ref{fig:TipVortex}.

\begin{figure}
\centering
\includegraphics[width=1\textwidth]{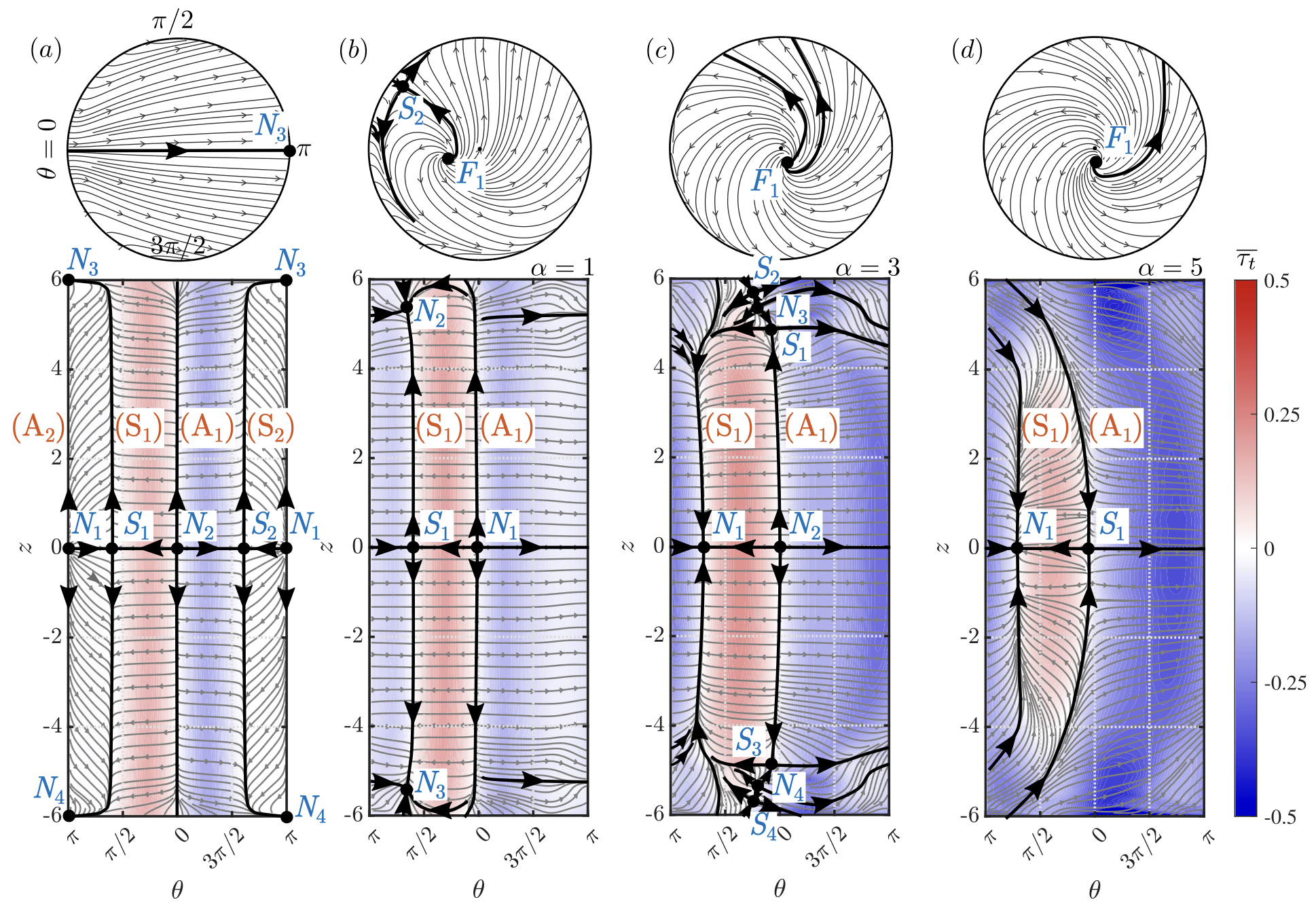}
\caption{Time-averaged skin friction line patterns for selected rotation rates with $AR=12$. $(a)$ $\alpha=0$, $(b)$ $\alpha=1$, $(c)$ $\alpha=3$ and $(d)$ $\alpha=5$. The top circles show the skin friction lines on the side surface of the cylinder, and the bottom figures show the skin friction lines on the $\theta$--$z$ plane. The color represents the skin friction vector in the circumferential direction $\overline{\tau_t}$. The critical points and the separation/attachment lines (S)/(A) are indicated in the figures.}
\label{fig:SFL}
\end{figure}

To provide a preliminary assessment of the free-end effects on the inboard flows, we present the skin friction line patterns for the cases with $AR=12$ in figure \ref{fig:SFL}. 
The critical-point theory \citep{Lighthill1963,Tobak1982,Delery2001} is also applied to construct a consistent flow topology on the rotating cylinder surface. 
The topological rules and critical point counts are examined, with the numbering of critical points in the surface flow patterns summarized in table \ref{table:critical_point} for different rotation rates. 
The number of critical points is verified to satisfy the Poincar\'e–Bendixson theorem \citep{Davey1961}, which dictates that the topological constraint $\sum N - \sum S =2$ must be maintained for fluid flows (here, $\sum N$ and $\sum S$ are the numbers of nodal points (including nodes and foci) and saddle points, respectively).

For the stationary cylinder, the boundary layer separation exhibits symmetry about both $\theta=0$ and $z=0$. 
The separation lines (S$_1$) and (S$_2$) pass through saddle points $S_1$ and $S_2$ shown in figure \ref{fig:SFL}($a$). 
While the boundary layer flow upstream of separation exhibits an almost two-dimensional pattern, the skin-friction lines downstream of separation display a strong spanwise flow component. 
This indicates that the wake of the stationary cylinder is significantly influenced by inboard flow from the free ends, a well-documented feature in free-end cylinder studies \citep{sumner2013flow,cao2022jfm}.

For the rotating cylinder, the surface friction line represents the relative motion of the boundary layer to the rotating surface.
In these cases, the separation line on the advancing side of the cylinder disappears. 
The boundary layer flow is accelerated between $\theta\approx 0$ to $\pi/2$, as indicated by the positive $\overline{\tau_z}$ regions in figure \ref{fig:SFL}($b$--$d$). 
This acceleration, by Bernoulli's theorem, induces a low-pressure region that is pivotal for generating lift.

At $\alpha=1$, a node $N_2$ forms on the retreating side, with its eigenvector line connecting to the separation line originating from saddle point $S_2$ on the free-end surface. 
The unstable focus $F_1$ dominates the spiral-out flow on the free-end surface, which feeds into saddle point $S_2$. 
On the cylinder surface, both the boundary layer flow in the acceleration and deceleration regions are mostly two-dimensional, except for the vicinity of the free ends.
This observation indicates that the formation of tip vortices with mild rotation alleviates the highly three-dimensional free-end effects observed in the case of a finite stationary cylinder.

As the rotation rate increases to $\alpha=3$, the saddle point $S_2$ shifts from the free-end boundary to the cylinder surface.
The attachment line (A$_1$) connects multiple critical points, forming the characteristic $N_2$--$S_1$--$N_3$--$S_2$ combination typical of reattachment flow patterns, a result of the highly three-dimensional flow structures near the free ends.
The free-end effects further intensifies as the rotation rate increases to $\alpha=5$, with the reattachment line (A$_1$) exhibiting greater curvature across a substantially wider span. 
Such strong three-dimensionality is likely linked with the inboard-wise tilt of the tip vortices at high rotation rates, as discussed in figure \ref{fig:TipVortex}.
As a result, the region associated with positive $\overline{\tau_t}$ is significantly reduced, leading to degraded aerodynamic performance.
These results demonstrate the significant impact of the free ends on the boundary flow topology, particularly for cases with high rotation rates.
In what follows, we discuss the three-dimensional wake dynamics of finite rotating cylinders, revealing their strong dependence on tip vortices.

\begin{table}
\setlength{\tabcolsep}{10pt}
 \begin{center}
  \begin{tabular}{lccccc}
    rotation rate  & Nodes $\sum N$   &   Foci $\sum F$ & Saddle points $\sum S$ \\[3pt]
       $\alpha =0$   & 4 & 0 & 2\\
       $\alpha =1$   & 3 & 2 & 3\\
       $\alpha =3$   & 4 & 2 & 4 \\
       $\alpha =5$   & 1 & 2 & 1 \\
  \end{tabular}
  \caption{
  Number of critical points in the time-averaged surface flow patterns with $AR=12$.}
  \label{table:critical_point}
 \end{center}
\end{table}

\subsection{Wakes of finite rotating cylinders}
\label{sec:overview}

\begin{figure}
\centering
\includegraphics[width=0.8\textwidth]{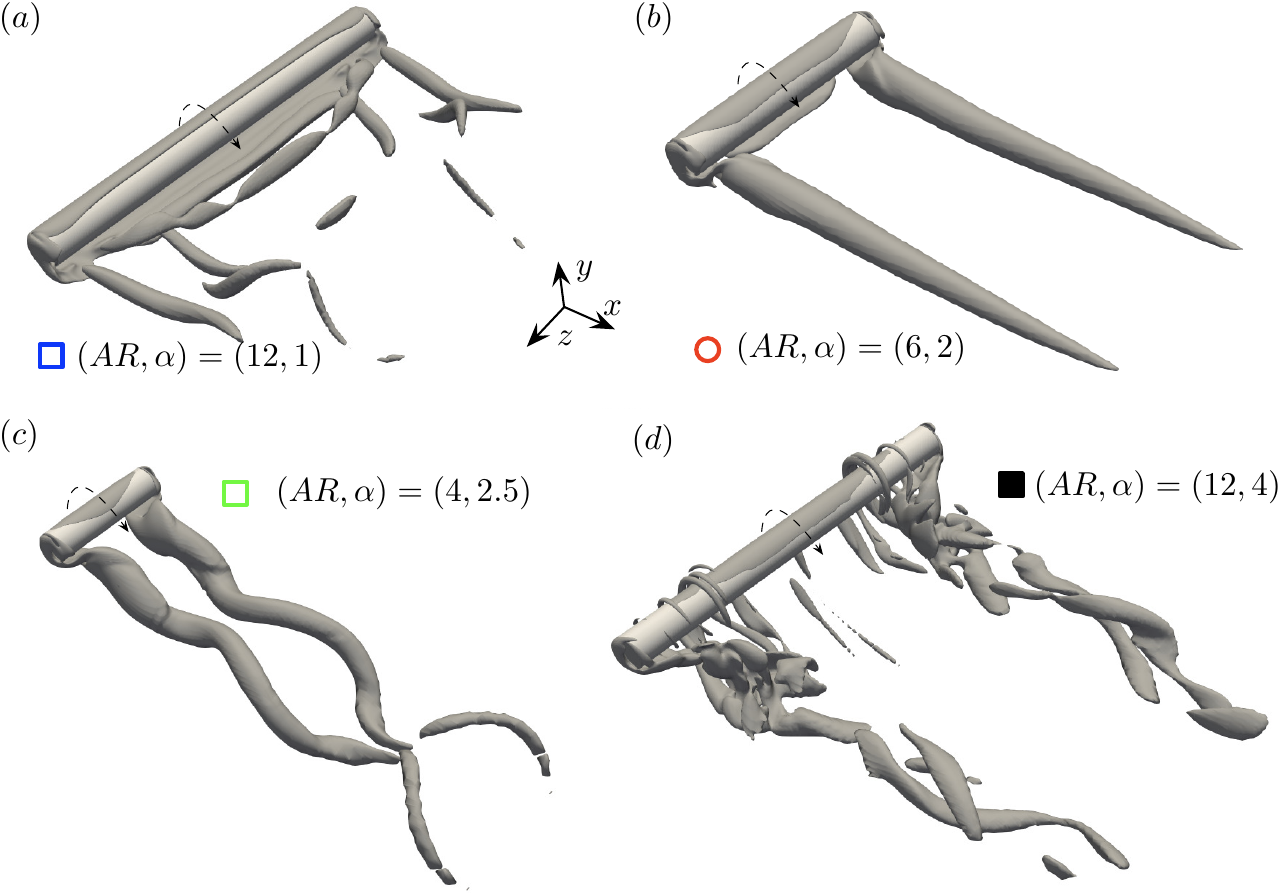}
\caption{Perspective views of the four typical wakes behind the rotating finite cylinders. ($a$) Unsteady vortex shedding, ($b$) steady flow, ($c$) unsteady tip vortices and ($d$) chaotic flow. The vortical structures are visualized by iso-surfaces of $Q=1$. The dashed arrows indicate the direction of rotation.}
\label{fig:FourTypes}
\end{figure}

\begin{figure}
\centering
\includegraphics[width=1\textwidth]{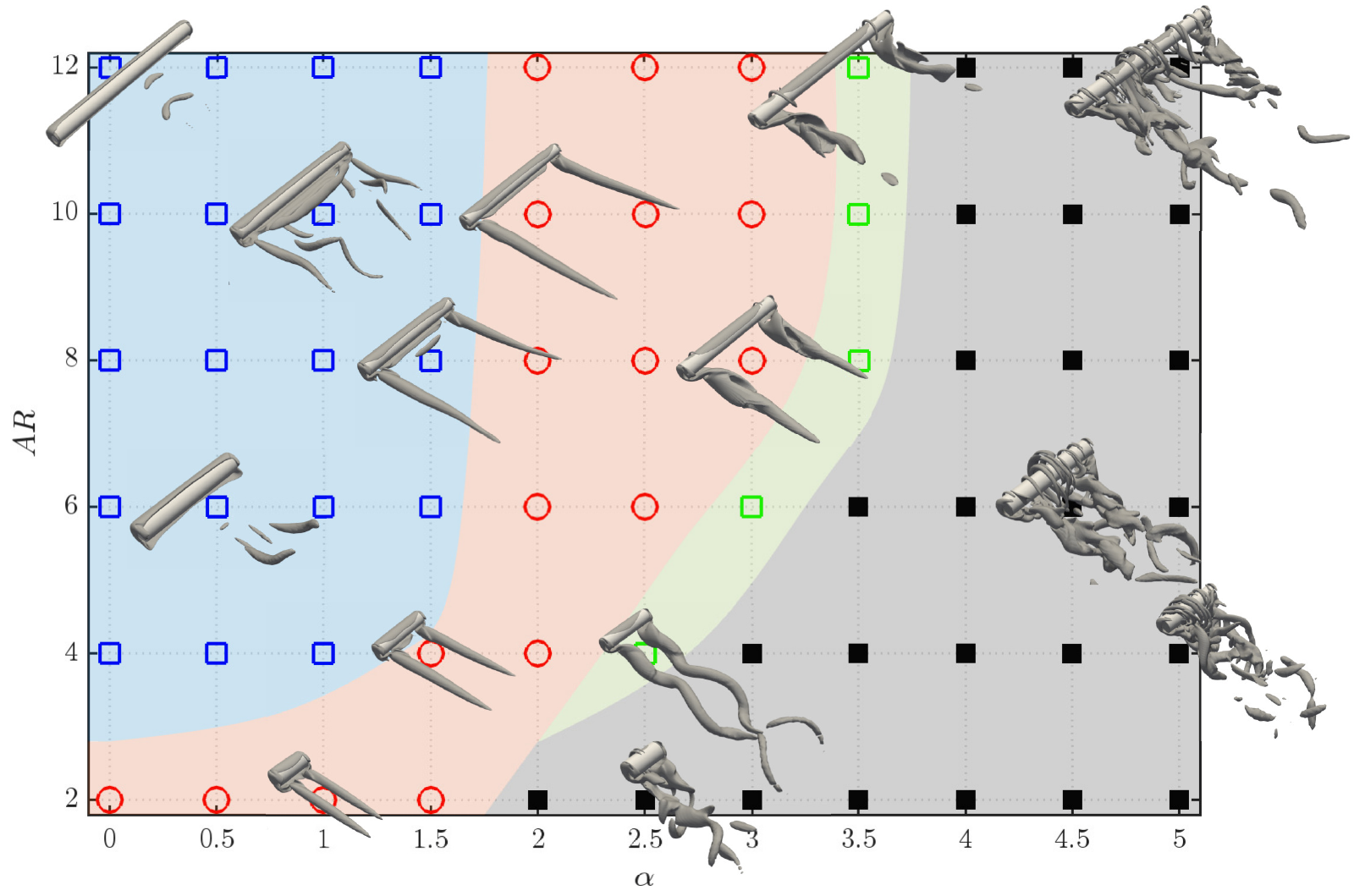}
\caption{Classification of wake regimes. Blue squares: unsteady vortex shedding along midspan; red circle: steady flows; green squares: unsteady tip vortices; black filled squares: chaotic flow.}
\label{fig:Regime}
\end{figure}

Under the influence of the tip vortices, we identify four types of wakes of the rotating cylinders at $Re=150$, including unsteady vortex shedding, steady flow, unsteady tip vortices, and chaotic flows, as shown in figure \ref{fig:FourTypes}.
The distributions of these flows in the $\alpha$--$AR$ space are mapped out in figure \ref{fig:Regime}. 
In the following, we discuss the characteristics of each type of flows.

\subsubsection{Unsteady vortex shedding}
\label{sec:UnsteadyVortexShedding}
At low rotation rates, the cylinder displays typical bluff body behavior, producing unsteady vortex shedding in its wake, as illustrated in figure \ref{fig:FourTypes}($a$) for the case $(AR,\alpha)=(12,1)$. 
The shed vortices are spatially confined to the midspan region due to the finite length of the cylinder. 
Adjacent to the midspan, the vortical structures are predominantly aligned along the spanwise direction.
Near the free ends, however, the shed vortices reorient themselves to align with the streamwise direction. These streamwise-oriented vortical structures, combined with the spanwise-oriented vortices near the midspan, form closed vortex loops that convect downstream.
Such three-dimensional vortex shedding pattern has been extensively documented in the wakes of stationary finite cylinders \citep{taneda1952studies,levold2012viscous} and in separated flows over finite wings \citep{zhang2020formation}.
In addition, the tip vortices exhibit unsteadiness under the influence of the unsteady vortex shedding.

\begin{figure}
\centering
\includegraphics[width=0.9\textwidth]{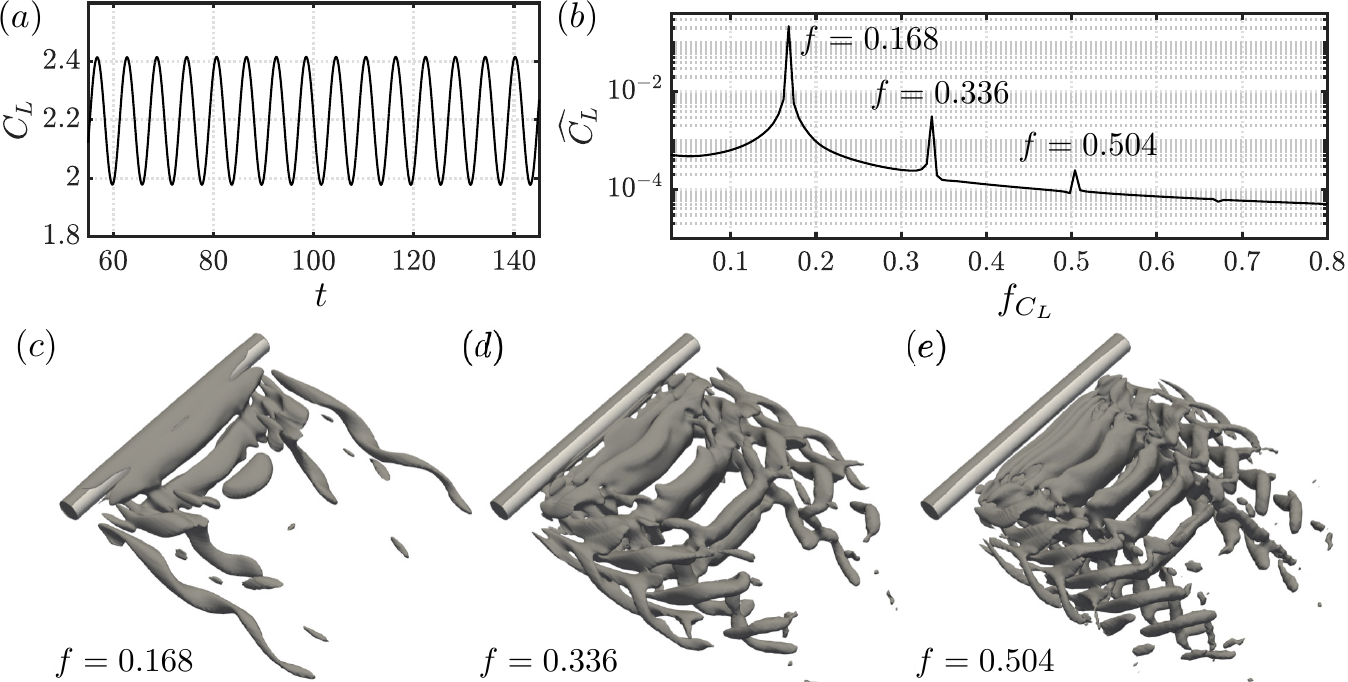}
\caption{Characterization of the unsteady vortex shedding for the case with $(AR,\alpha)=(12,1)$. ($a$) Time history of the lift coefficient, ($b$) amplitude spectrum of the lift coefficient, ($c$)--$(e)$ dynamic modes associated with the dominant frequency and its two superharmonics. The modal structures are visualized by isosurfaces of $Q$. }
\label{fig:AR12Omega2DMD}
\end{figure}

Despite the complex vortex shedding pattern described above, the lift coefficient for the case $(AR, \alpha) = (12, 1)$ exhibits a regular periodic oscillation dominated by a single frequency at $f = 0.168$, as seen in figure \ref{fig:AR12Omega2DMD}($a, b$).
Additionally, two superharmonic frequencies at $2f$ and $3f$ emerge due to the nonlinear self-interaction of the dominant frequency.
To examine the spatial structures corresponding to these frequencies, we perform dynamic mode decomposition (DMD) \citep{SCHMID_2010} using the snapshots of $Q$, with the resulting modes displayed in figure \ref{fig:AR12Omega2DMD}($c$--$e$).
The dominant frequency corresponds to a highly three-dimensional modal structure, consisting of spanwise vortices in the midspan near-wake region, streamwise-oriented vortices, and helical tip vortices. 
The modal structures of the superharmonics feature smaller-scale vortices that extend further downstream into the wake.

\begin{figure}
\centering
\includegraphics[width=1\textwidth]{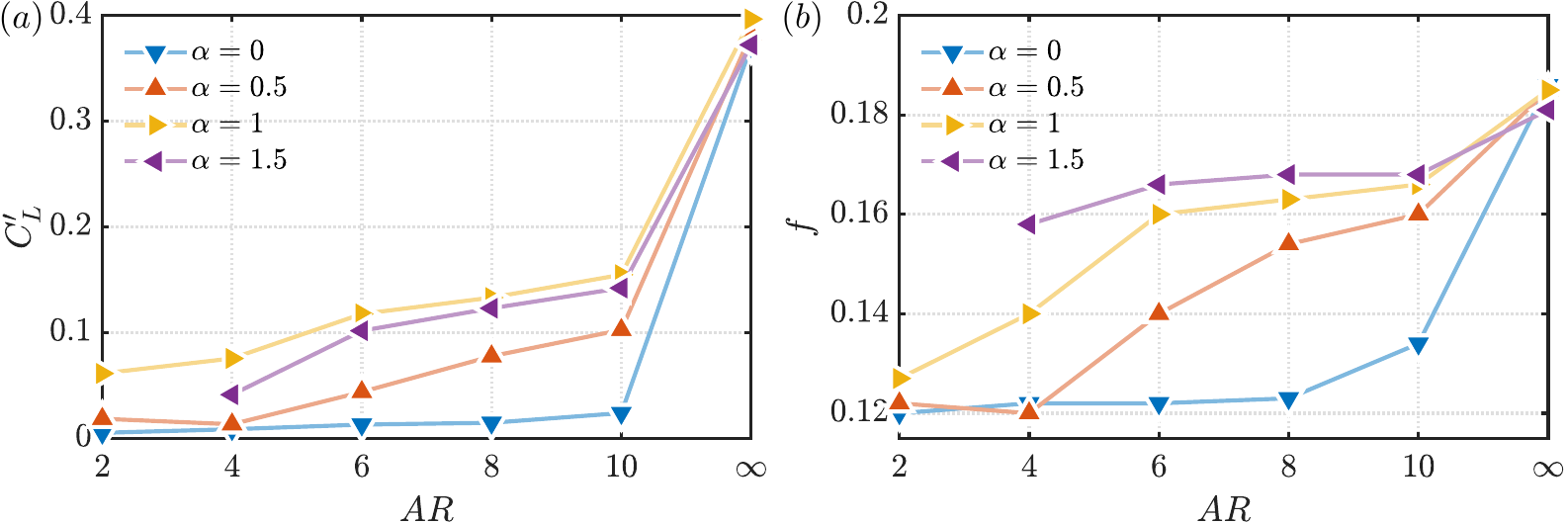}
\caption{Compilation of ($a$) root mean squared value of the fluctuating lift coefficients  and $(b)$ dominant frequency for the cases in the unsteady vortex shedding regime.}
\label{fig:UnsteadyCompile}
\end{figure}

The effects of the aspect ratio on the fluctuating lift coefficient and dominant vortex shedding frequency are summarized in figure \ref{fig:UnsteadyCompile}.
While in two-dimensional flows around rotating cylinders (i.e., $AR=\infty$) the rotation rate $\alpha$ has a limited effect on vortex shedding strength and frequency until the flow transitions to a steady wake \citep{mittal2003flow}, the behavior of finite rotating cylinders differs significantly.
The root-mean-square (r.m.s.) values of lift coefficients are substantially higher for rotating finite cylinders compared to their static peers, suggesting that mild rotation enhances vortex shedding. 
This distinction arises from fundamentally different end effects between static and rotating configurations. 
In the former, the free-end effects primarily manifest as a inboard-wise flow stream from the free end towards the near wake of the midspan region\footnote{Although this end-to-middle-span flow is commonly termed ``downwash'' in studies of wall-mounted cylinders ($z$ direction), we avoid this terminology to prevent confusion with the tip-vortex-induced downwash, which points in the $-y$ direction.}, as discussed in figure \ref{fig:SFL}($a$).
This spanwise flow acts as a splitter plate in the near wake and significantly hinders the formation of a K\'arm\'an vortex street \citep{KRAJNOVIC_2011,WANG_ZHOU_2009,cao2022jfm}.
In contrast, rotating cylinders generate the streamwise-oriented tip vortices that induce downwash (i.e., velocity component in the $-y$ direction). 
While this free-end effects significantly alter the vortex shedding behavior as described in figure \ref{fig:AR12Omega2DMD}(c--e), the strength of vortex shedding remains strong compared to the static cylinder.
However, as aspect ratio decreases, the intensified downwash increasingly suppresses the vortex shedding process, reducing both its frequency and strength.
At sufficiently low aspect ratios, e.g., $(AR,\alpha)=(4,1.5)$, the downwash effects lead to complete wake stabilization, which we examine in detail in the next subsection.

\subsubsection{Steady flow}
\label{sec:SteadyWake}
For rotation rates $\alpha \gtrsim 2$, the increased angular velocity confines vorticity to the near-wall region, thickening the zone of closed streamlines. This confinement significantly restricts vorticity transport into the wake, suppressing separation bubble formation and inhibiting the development of wake instabilities \citep{mittal2003flow,rao2015review}. As a result, the flow transitions to a steady state, as illustrated in figure \ref{fig:FourTypes}$(b)$.

\begin{figure}
\centering
\includegraphics[width=0.9\textwidth]{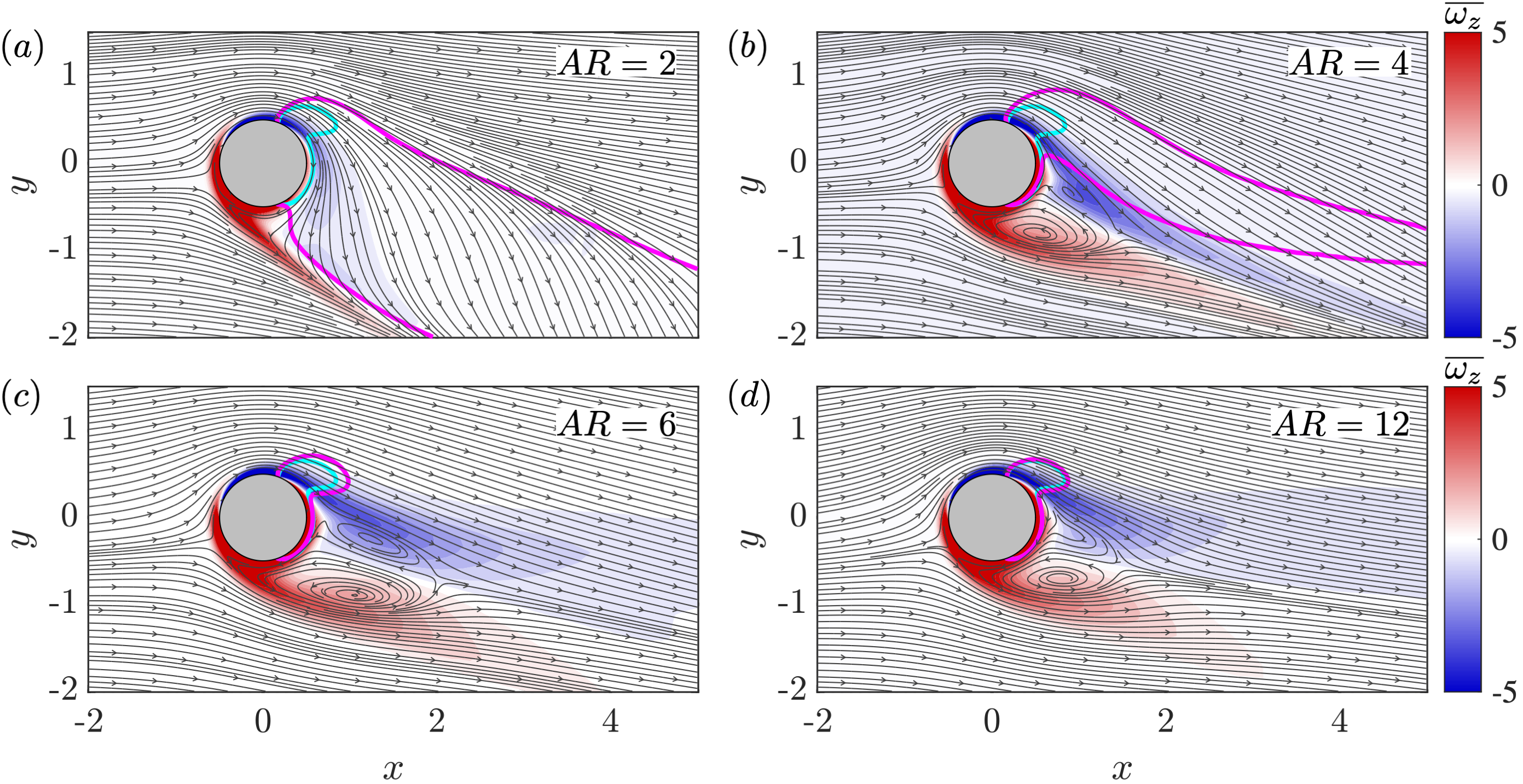}
\caption{Time-averaged streamlines and the vorticity fields at the middle section $(z=0)$ of of cases with different aspect ratios at a fixed spin rate $\alpha=1.5$. The magenta and cyan lines represent the contours of $\overline{U_y}=-0.5$ for the three-dimensional case and the 2D case, respectively.}
\label{fig:Omega3MidSpan}
\end{figure}

Steady flow can also be observed for $\alpha \lesssim 2$ with low-aspect-ratio rotating cylinders, where wake stabilization is achieved by the downwash effects from the tip vortices. To demonstrate this mechanism, we examine cases with $\alpha = 1.5$, in which the analogous two-dimensional flow exhibits classic vortex shedding.
Figure \ref{fig:Omega3MidSpan} presents time-averaged flow fields at the midspan ($z=0$) for $\alpha = 1.5$ across varying $AR$. Contours of $\overline{U_y} = -0.5$ are included for both finite-$AR$ (magenta) and two-dimensional (cyan) cases to highlight the influence of tip vortices on the wake velocity distribution.
For $AR \gtrsim 6$, the two contours nearly coincide, indicating minimal downwash effects at the midspan. Consequently, the wake remains unsteady, with vortex shedding persisting near the midspan and forming a pair of separation bubbles, as shown in figures \ref{fig:Omega3MidSpan}$(c,d)$.
When the aspect ratio is reduced to $AR=4$, downwash effects intensify, evidenced by the expanded region occupied by the $\overline{U_y} = -0.5$ contour. This downward velocity impinges on the upper vortex bubble, suppressing its roll-up and stabilize the wake.
This mechanism is analogous to the confinement of leading-edge vortex (LEV) in low-aspect-ratio wings \citep{taira2009three,DeVoria_Mohseni_2017,zhang2020laminar,zhang2020formation}.
At even lower aspect ratios, tip vortices dominate the wake, and their strong downwash completely eliminates separation bubbles, as shown in figure \ref{fig:Omega3MidSpan}$(a)$.

\subsubsection{Unsteady tip vortices}
\label{sec:UnsteadyTipVortices}

\begin{figure}
\centering
\includegraphics[width=0.9\textwidth]{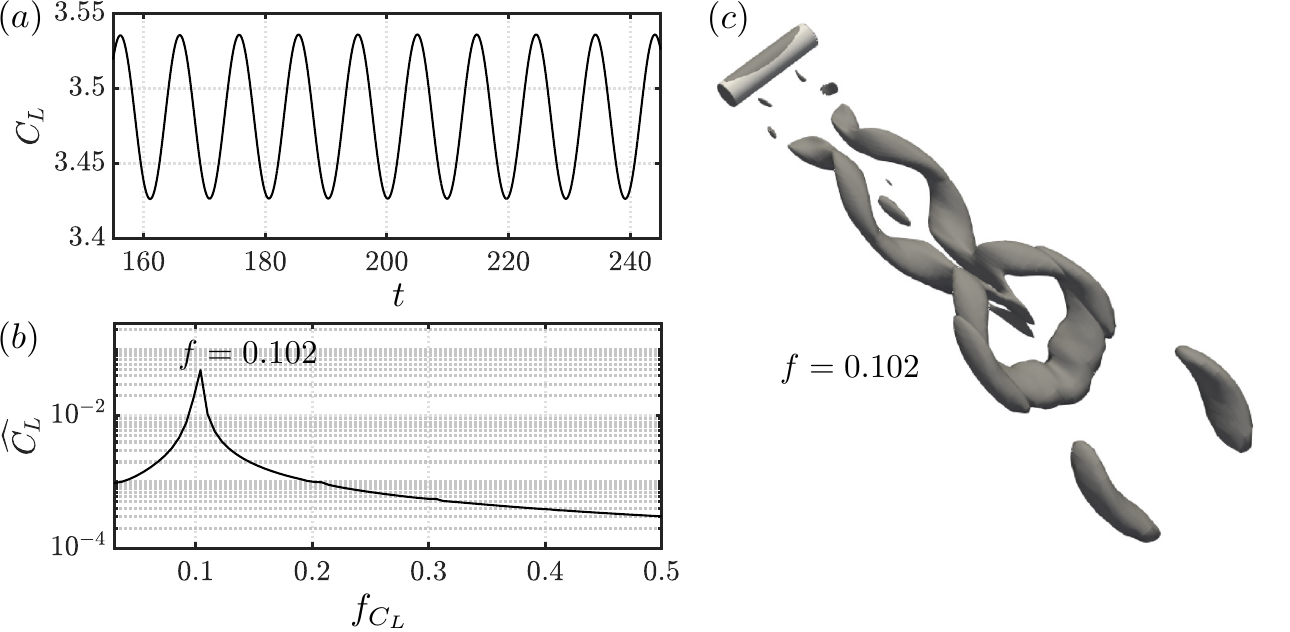}
\caption{Characteristics of the unsteady tip vortices for the case $(AR,\alpha)=(4,2.5)$. ($a$) Time history of the lift coefficient, ($b$) amplitude spectrum of the lift coefficient, ($c$) DMD mode at $f=0.102$. }
\label{fig:AR4Omega5DMD}
\end{figure}

As the rotation rate increases, the steady wake becomes unstable due to disturbances originating in the tip vortices. The unsteadiness arises from two primary mechanisms: the mutual induction between counter-rotating tip vortex pairs or the inherent instability of an individual vortex as its strength intensifies. These mechanisms align with the tip vortex behavior described in \S \ref{sec:tipVortex} for the unsteady tip vortex regime.
The mutual induction mechanism is prevalent in cases with low aspect ratios, such as $(AR, \alpha) = (4, 2.5)$ depicted in figure \ref{fig:FourTypes}($c$). Spectral analysis of this flow, shown in figure \ref{fig:AR4Omega5DMD}, reveals a nearly pure sinusoidal oscillation, as the lift coefficient is dominated by a single frequency with negligible harmonics. 
The corresponding DMD mode illustrates a pair of serpentine tip vortices oscillating symmetrically in a varicose pattern. 
These vortices eventually merge downstream to form closed vortex rings.
For this instability mode, the critical rotation rate $\alpha$ scales nearly linearly with aspect ratio for $AR \lesssim 8$. 
This trend reflects the need for greater vortex intensity to induce instability via mutual induction as the spanwise separation between counter-rotating vortices increases. 
Beyond $\alpha \approx 3.5$, further increases in rotation rate transition the flow to a chaotic state, which will be addressed in \S \ref{sec:Taylor-likeVortex}.

\begin{figure}
\centering
\includegraphics[width=0.85\textwidth]{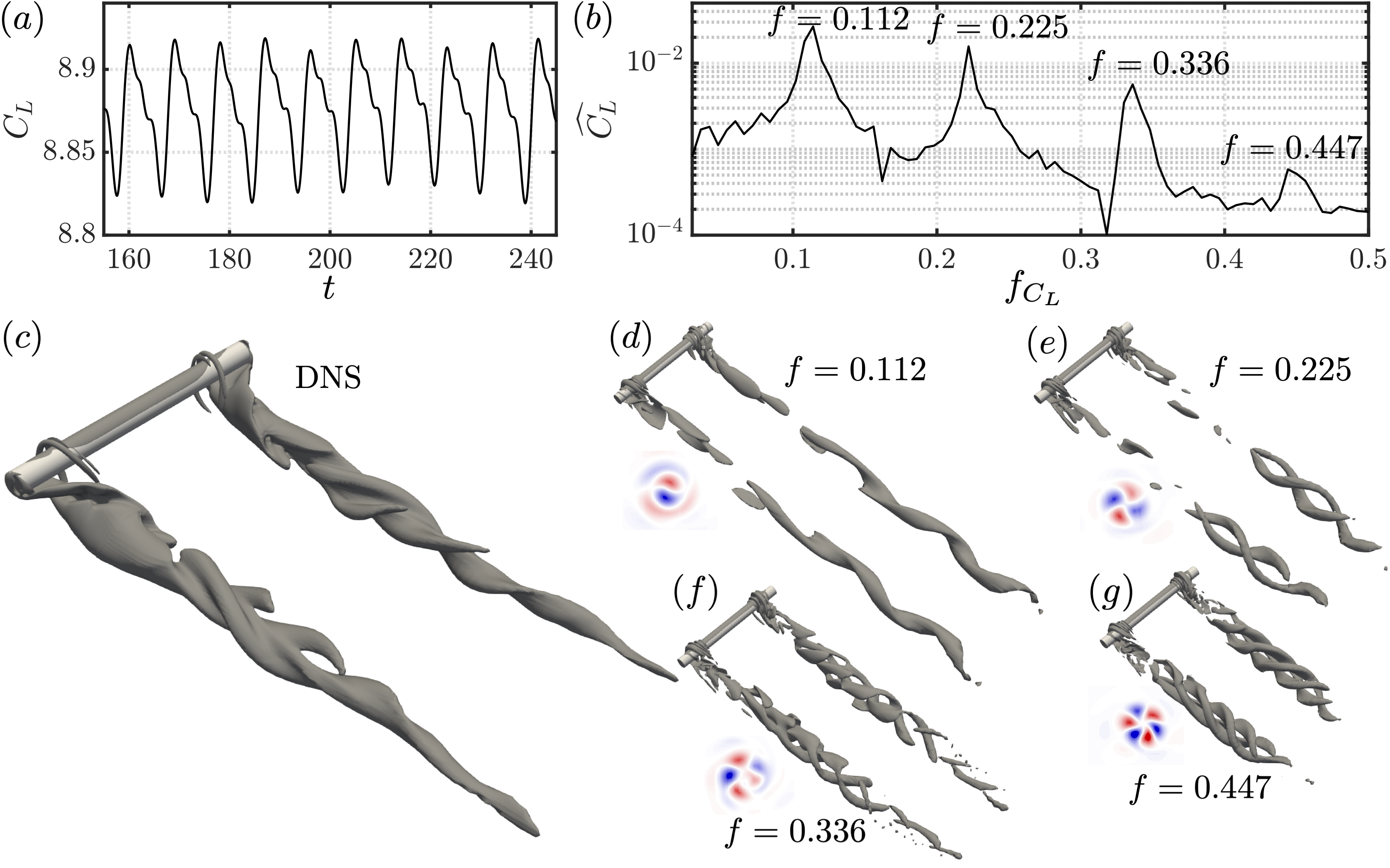}
\caption{Characteristics of the unsteady tip vortices for the case $(AR,\alpha)=(12,3.5)$. ($a$) Time history of the lift coefficient, ($b$) amplitude spectrum of the lift coefficient, ($c$) a snapshot of flow field visualized by isosurfaces of $Q=0.5$. $(d)$--$(g)$ DMD modes corresponding to the dominant frequencies. A slice of the modal structure on the $y-z$ plane at $x=20$ is added as inset for each DMD mode, with red and blue representing $Q>0$ (vortex core) and $Q<0$ (strain region), respectively.}
\label{fig:AR12Omega7DMD}
\end{figure}

For cases with high aspect ratios, the intensification of tip vortices can trigger another type of unsteadiness as illustrated in figure \ref{fig:AR12Omega7DMD} for the case $(AR,\alpha)=(12,3.5)$.
In this case, the lift coefficient displays fluctuations governed by a fundamental frequency of $f=0.112$, accompanied by superharmonics of comparable magnitude.
The flow field exhibits distinct characteristics across different wake regions. 
In the near wake, the vortex cores exhibit irregular geometries. 
Downstream in the far wake, these cores evolve into corkscrew-like configurations characterized by secondary vortex filaments that interweave along the primary cores.
DMD analysis reveals that the far wake fluctuations feature a single spiral mode ($m=1$) for the dominant frequency, while the $2f$ and $4f$ are associated with double ($m=2$) and triple ($m=3$) braided helix modes. 
The $3f$ DMD mode exhibits a mixed structure, combining characteristics of both the $m=2$ and $m=3$ helix modes—likely due to nonlinear interactions in the wake.

\subsubsection{The Taylor-like vortices}
\label{sec:Taylor-likeVortex}

\begin{figure}
\centering
\includegraphics[width=0.85\textwidth]{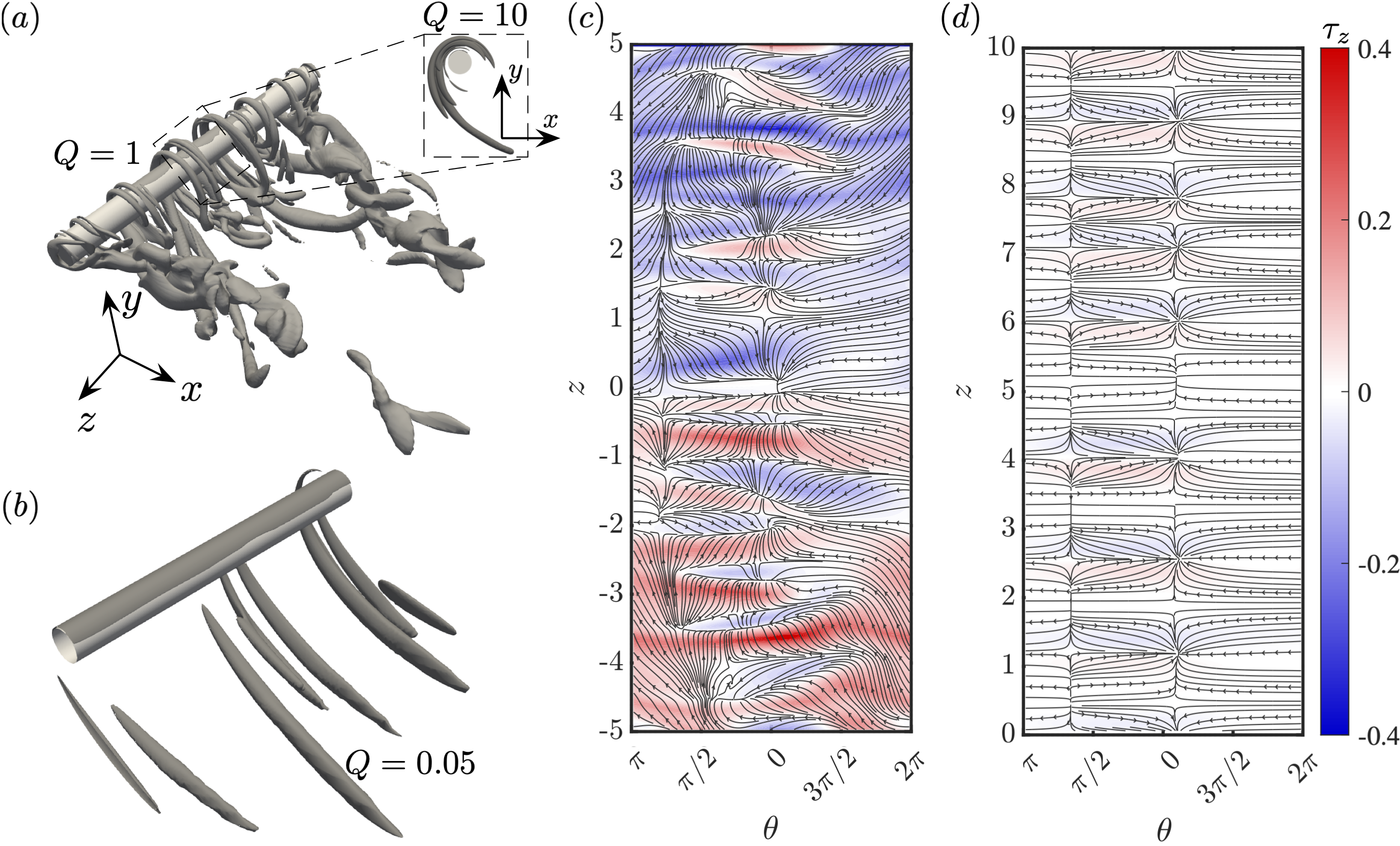}
\caption{Comparison of the vortical structures between the ($a$) free-end case $(AR, \alpha)=(10, 5)$ and ($b$) spanwise-periodic case at $\alpha=5$. In the former, the vortical structures are visualized by isosurfaces of $Q=1$, and latter $Q=0.05$. In the inset plot in ($a$), $Q=10$ is used only near the midspan of the cylinder to show the shape of the Taylor-like vortices. $(c)$ and $(d)$ show the instantaneous skin friction line patterns for the two cases, colored by the spanwise component of the skin friction vector $\tau_z$.}
\label{fig:CompareWithSlipQ}
\end{figure}

At even higher rotation rates, the three-dimensional wake undergoes a transition to chaotic flow. For low-aspect-ratio configurations, this transition occurs at lower $\alpha$, primarily due to strong interactions between counter-rotating tip vortices. 
In contrast, for high-aspect-ratio cases, the onset of chaotic flow is dominated by the formation of Taylor-like vortices (analogous to the toroidal vortices in the Taylor-Couette flow observed between concentric rotating cylinders). 
This subsection focuses on the latter mechanism.

The Taylor-like vortices typically emerge for $\alpha \gtrsim 4$ in the case of finite rotating cylinders, as displayed in figure \ref{fig:CompareWithSlipQ}($a$) for the case $(AR, \alpha) = (10, 5)$.
These vortices appear as C-shaped arc structures that are bound to the windward and retreating sides of the cylinder surface, while extending further away from the advancing side and the wake.
Some of the vortical structures are absorbed into the tip vortex region, as corroborated by the discussion regarding figure \ref{fig:TipVortex}.
The Taylor-like vortices are populated along the cylinder span, imprinting intricate skin-friction line patterns on the surface of the cylinder, as shown in figure \ref{fig:CompareWithSlipQ}$(c)$.
In contrast, an infinitely long rotating cylinder with analogous condition exhibits the typical mode E instability as shown in figure \ref{fig:CompareWithSlipQ}($b$).
In this case, the flow's three-dimensionality is so weak that a significantly lower contour level of $Q$ is required to reveal the wake structures. 
The wake is characterized by pairs of counter-rotating streamwise vortices on the bottom side of the cylinder, which form regular cell-like patterns in the skin-friction lines depicted in figure \ref{fig:CompareWithSlipQ}($c$).
This comparison between the free-end and two-dimensional cases highlights that the highly three-dimensional wake dynamics in the former case are primarily attributed to the extrinsic free-end effects, rather than being intrinsically driven.

\begin{figure}
\centering
\includegraphics[width=1\textwidth]{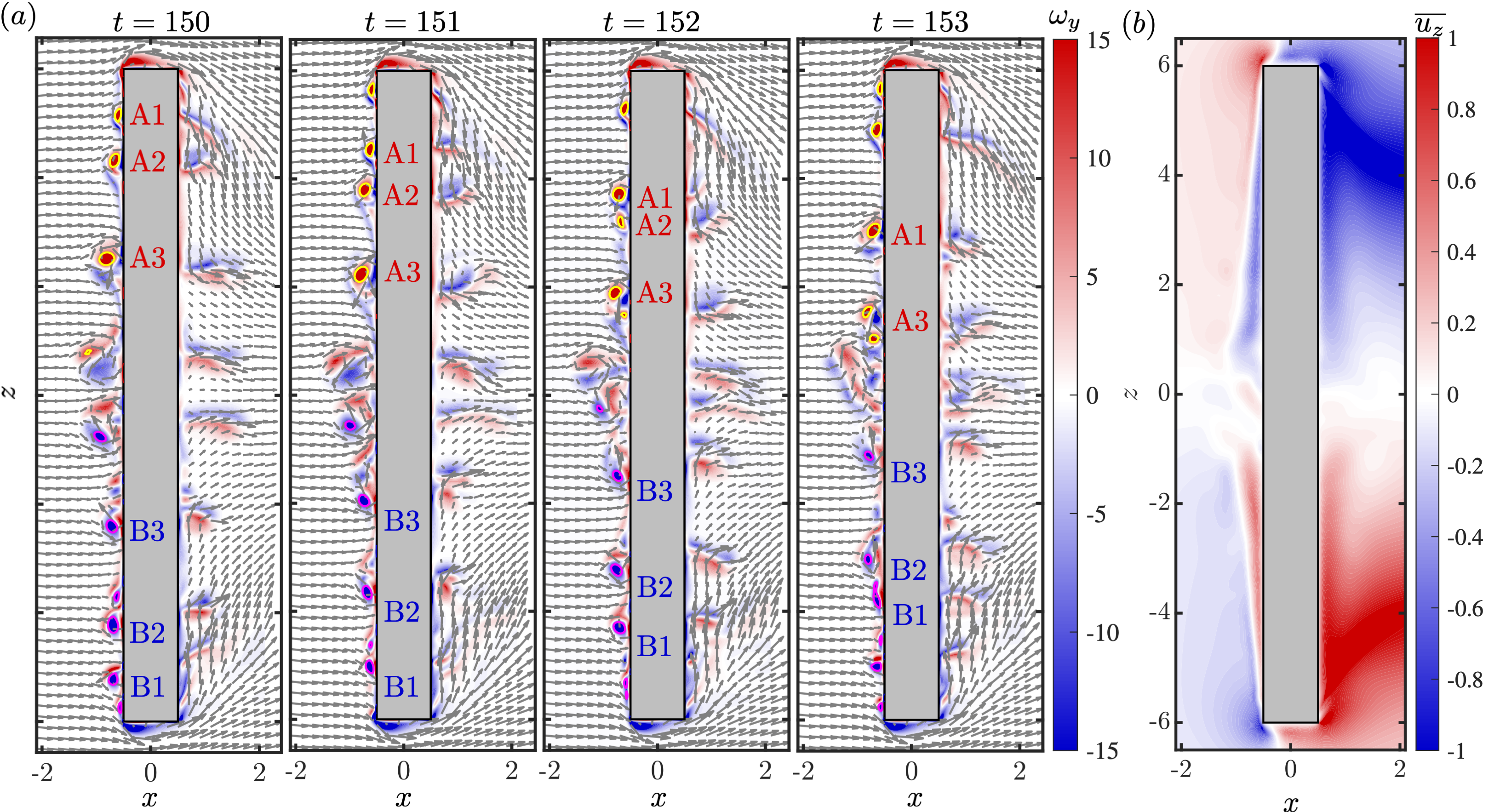}
\caption{$(a)$ Temporal evolution of the instantaneous $\omega_y$ on the $y=0$ plane at $t=150 - 153$ for $(AR,\alpha)=(12,5)$. Velocity vectors are overlaid as quiver plots. The Taylor-like vortices are indicated by yellow ($\omega_y=15$) and cyan $(\omega_y=-15)$ contours, and are labeled as A1-A3 and B1-B3, respectively. ($b$) Time-averaged spanwise velocity $\overline{u_z}$ on the $y=0$ plane.}
\label{fig:OmegaYEvolution}
\end{figure}

\begin{figure}
\centering
\includegraphics[width=0.8\textwidth]{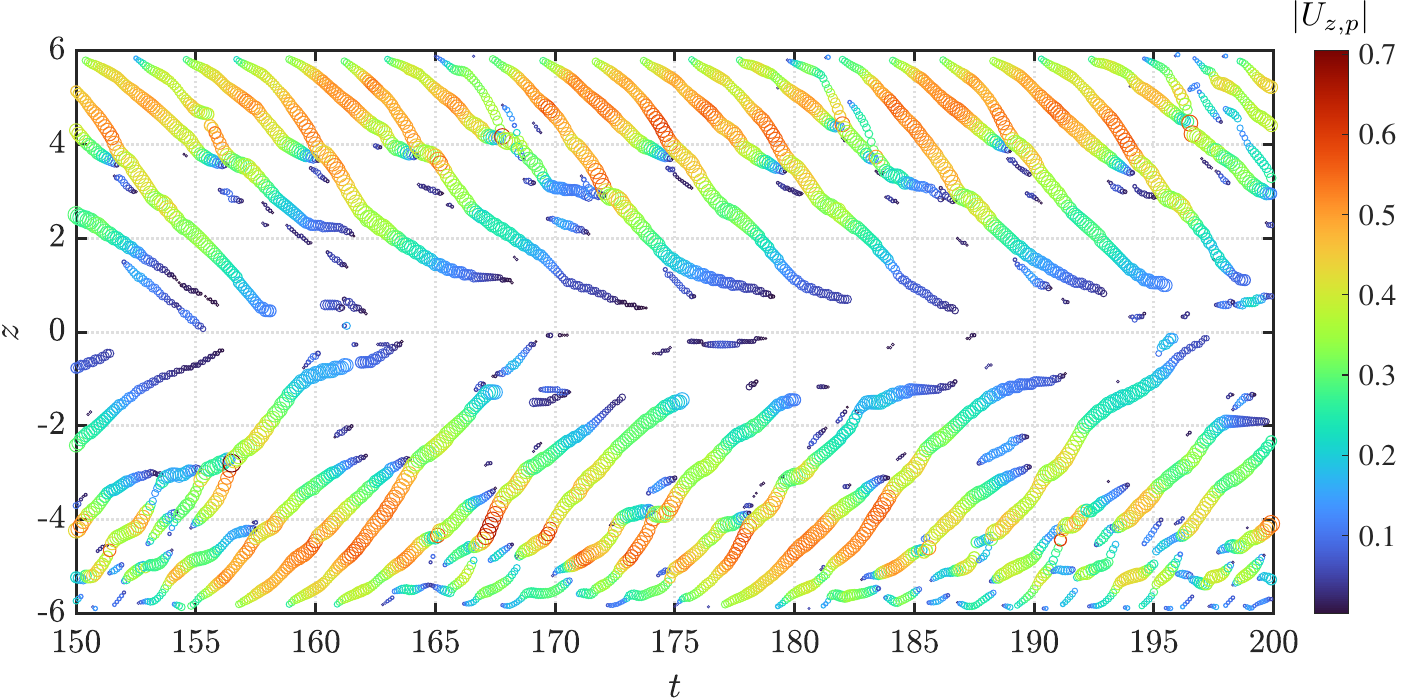}
\caption{Spatio-temporal distribution of the Taylor-like vortices for $(AR,\alpha)=(12,5)$ on the $y=0$ plane. Each circle denote a contour of $\omega_y=15$ for $z>0$ and $\omega_y=-15$ for $z<0$ on the plane of $y=0$. The size of the circle represent the circulation of the vortices. The color depicts the absolute value of the spanwise velocity predicted by the potential flow theory.}
\label{fig:TaylorDynamics}
\end{figure}

The evolution of the Taylor-like vortices on the plane of $y=0$ is illustrated in figure \ref{fig:OmegaYEvolution}($a$) for the case $(AR,\alpha)=(12,5)$ using four successive snapshots of $\omega_y$ fields.
A more complete picture of the spatio-temporal dynamics of these Taylor-like vortices is shown in figure \ref{fig:TaylorDynamics}, in which the occurrence of a Taylor-like vortex (defined as closed contour lines of $\omega_y=15$ for $z>0$ and $\omega_y=-15$ for $z<0$ as shown in figure \ref{fig:OmegaYEvolution}) in the $t$-$z$ diagram is denoted by a circle. 
Over time, the Taylor-like vortices consistently emerge from the free ends, and grow stronger as they migrate towards the midspan.
The windward side of the cylinder surface (where the Taylor-like vortices are bounded) is predominantly populated with positive $\omega_y$ for $z>0$, and negative $\omega_y$ for $z<0$. 
A notable feature is the frequent merging of adjacent vortices into single structures at approximately 2-3 diameters from the free ends.
This vortex coalescence phenomenon is clearly captured in the snapshot at $t=152$, where vortices A1 and A2 combine to form a unified Taylor-like vortex.
Eventually, the counter-rotating Taylor-like vortices developed from the two free ends collide near the midspan, leading to complex vortex interactions.
The dynamics of the Taylor-like vortices is also depicted in the movie in supplementary material, where isosurfaces of $Q=1$ is visualized.
The characteristics of the Taylor-like vortices described above bear resemblance to the mode F instability in the two-dimensional flows over rotating cylinder, although the latter mechanism typically occurs at higher Reynolds numbers through centrifugal instability \citep{Radi2013JFM,rao2015review}.
Due to the consistent inboard-wise movement of the Taylor-like vortices, strong spanwise velocity is observed in the close vicinity of the cylinder surface in the time-averaged flow fields, as shown in figure \ref{fig:OmegaYEvolution}.

We argue that the inboard movement of Taylor-like vortices is attributed to the self-induced motion arising from vortex-wall interaction.
In potential flow theory, when a vortex is generated near the cylinder surface, an imaginary ``image vortex'' of opposite circulation is conceptually placed on the mirror side of the noslip surface to satisfy the no-penetration condition at the wall. 
The wall-normal velocity components of the real and image vortices cancel out to meet the boundary condition, but their tangential components add together, creating a net flow parallel to the wall that propels the Taylor-like vortices \citep{milnethomson1968theoretical}.
The self-induced velocity via vortex-wall interaction is predicted by the method of image as 
\begin{equation}
    U_{z,p} = \frac{\Gamma_T}{4\pi d_T},
\end{equation}
where $\Gamma_T$ is the circulation of the Taylor-like vortex, $d_T$ is the distance of the vortex to the wall.
Here, we calculate the circulation using $\Gamma_T=\oint_{\omega_y=\pm 15} \boldsymbol{u} \cdot \mathrm{d} \boldsymbol{l}$, and $d_T$ is defined as the distance between the vorticity extremum in the vortex core and the cylinder surface.
Although this potential-flow-based velocity does not quantitatively match the actual spanwise velocity of the Taylor-like vortices, it exhibits excellent qualitative agreement with the observed velocity trends (indicated by the slope of vortex trajectories in the $t$-$z$ plane). 
This correspondence strongly suggests that the inboard migration of Taylor-like vortices is indeed governed by the vortex-wall interaction mechanism.

\begin{figure}
\centering
\includegraphics[width=1\textwidth]{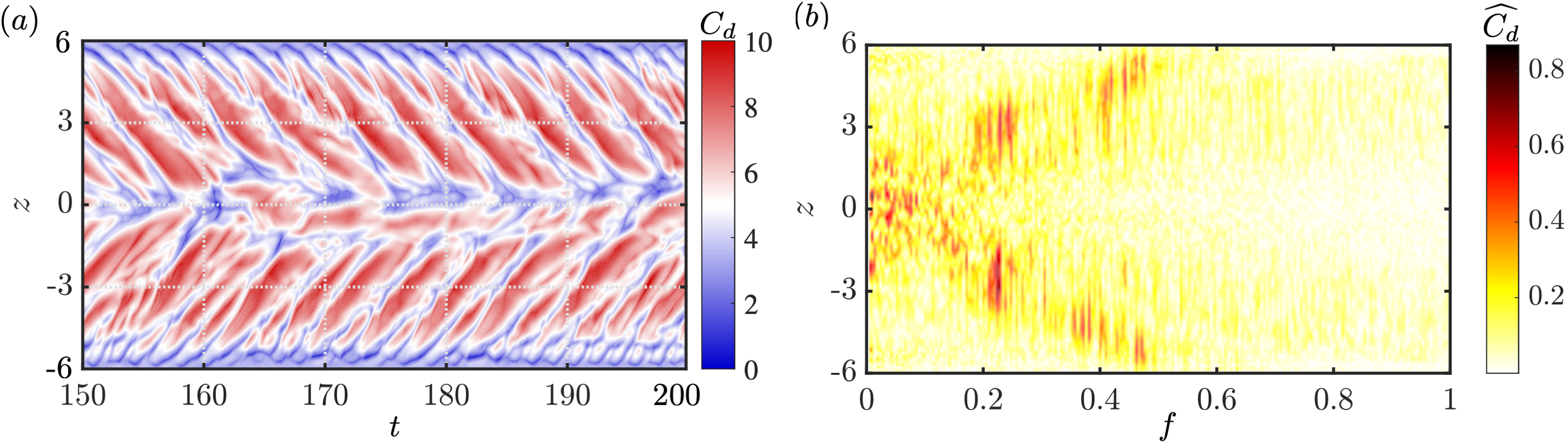}
\caption{ ($a$) Sectional drag coefficients for $(AR,\alpha)=(12,5)$ and ($b$) the amplitude spectrum.}
\label{fig:AR12Omega10SpatialTemporalSectionalForces}
\end{figure}

Since these Taylor-like vortices are spatially close to the windward side of the cylinder surface, they provide suction forces that reduce the local drag.
To demonstrate this, we present the distributions of the sectional drag coefficients along the spanwise direction and across time in figure \ref{fig:AR12Omega10SpatialTemporalSectionalForces}.
As the Taylor-like vortices propel toward the midspan, they induce slanted troughs of reduced drag in the spatio-temporal pattern of $C_d$. 
Notably, due to the accumulation of these vortices from both free ends, the drag coefficient at the midspan is relatively lower than the neighboring regions.
From the amplitude spectrum shown in figure \ref{fig:AR12Omega10SpatialTemporalSectionalForces}($b$), the regions near the free ends ($0\lesssim |z|\lesssim 3$) exhibit broadband frequency content centered around $f\approx 0.35-0.5$.
After the merge of the Taylor-like vortices, the region $2\lesssim |z| \lesssim 4$ is dominated by more concentrated frequencies within $f\approx 0.2-0.25$.
Toward the midspan, the vortices experience low-frequency broadband oscillations due to the complex interactions between Taylor-like vortices of opposite signs.

\subsection{Aerodynamic forces}
\label{sec:forces}

\begin{figure}
\centering
\includegraphics[width=1\textwidth]{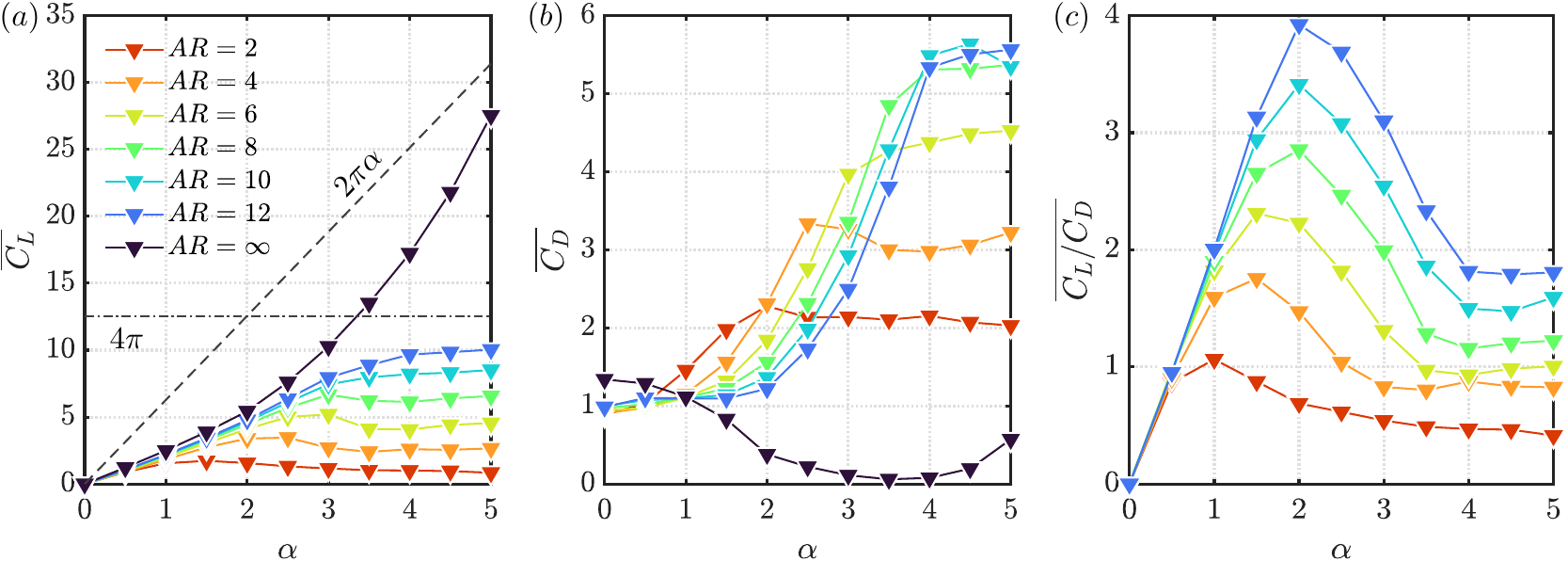}
\caption{Summary of aerodynamic forces. ($a$) Time-averaged lift coefficients, $(b)$ time-averaged drag coefficients and $(c)$ time-averaged lift-to-drag ratio. The potential flow solution ($C_L=2\pi\alpha$) and Prandtl's limit ($C_L=4\pi$) are included in ($a$).}
\label{fig:ForceCoeffs}
\end{figure}

The time-averaged lift and drag coefficients of the rotating cylinders are summarized in figure \ref{fig:ForceCoeffs}.
In the idealized two-dimensional flows ($AR = \infty$), the lift coefficient increases approximately quadratically with the rotation rate $\alpha$, exceeding Prandtl’s theoretical limit of $C_L = 4\pi$ \citep{prandtl1925magnuseffekt}, but remaining below the potential flow prediction of $C_L = 2\pi\alpha$ over the studied range of $\alpha$.
The drag coefficient, on the other hand, decreases with increasing $\alpha$ up to approximately $\alpha \approx 3.5$, approaching nearly zero, and then increases slightly at higher rotation rates. 

For finite rotating cylinders, the aspect ratio exerts a positive influence on the lift coefficient $C_L$, with higher $AR$ generally leading to increased lift, as shown in figure \ref{fig:ForceCoeffs}($a$).
However, even at large $AR$, the maximum $C_L$ remains significantly lower than that of the idealized two-dimensional case due to the presence of three-dimensional effects near the free ends. 
At low to moderate rotation rates, the lift increases with $\alpha$ as a result of the enhanced Magnus effect generated by surface rotation. 
As $\alpha$ increases further, the lift coefficient tends to plateau, and may even decline, due to the enhanced three-dimensional effects. 
The onset of this plateauing behavior is delayed for cylinders with larger aspect ratios, suggesting a greater ability to sustain two-dimensional-like lift generation over a broader range of rotation rates.
%Notably, for these finite-$AR$ cases, the lift coefficient does not exceed Prandtl's theoretical limit.

The drag characteristics of finite rotating cylinders differ significantly from those of the idealized two-dimensional case.
At low to moderate rotation rates, the drag coefficient initially increases with $\alpha$. 
However, this increase reaches a saturation point at a slightly higher rotation rate than the onset of lift plateauing.
The influence of aspect ratio on drag is also rotation-rate dependent. 
At low $\alpha$, drag decreases with increasing $AR$, whereas at high rotation rates, the drag becomes positively correlated with $AR$, likely due to stronger three-dimensional effects such as intensified tip vortices. 
Beyond $AR \approx 8$, however, the drag coefficients of the high-$\alpha$ cases also appear to saturate with increasing $AR$.

As a key measure of aerodynamic efficiency, the lift-to-drag ratio ($\overline{C_L/C_D}$) demonstrates a consistent monotonic increase with aspect ratio across all rotation rates, as shown in figure \ref{fig:ForceCoeffs}($c$).  
With increasing $AR$, the rotation rate at which peak $\overline{C_L/C_D}$ is achieved shifts from $\alpha=1$ to 2. 
Compared to finite wings under analogous flow conditions investigated in our previous study \citep{zhang2020formation}, the finite rotating cylinders generate substantially higher lift.
Although they also experience greater drag, the superior lift production results in an improvement in aerodynamic efficiency over conventional finite wings.
This advantage highlights the potential of rotating cylinders as a viable alternative to conventional lifting surfaces in applications where low-$Re$ performance is critical.

\begin{figure}
\centering
\includegraphics[width=1\textwidth]{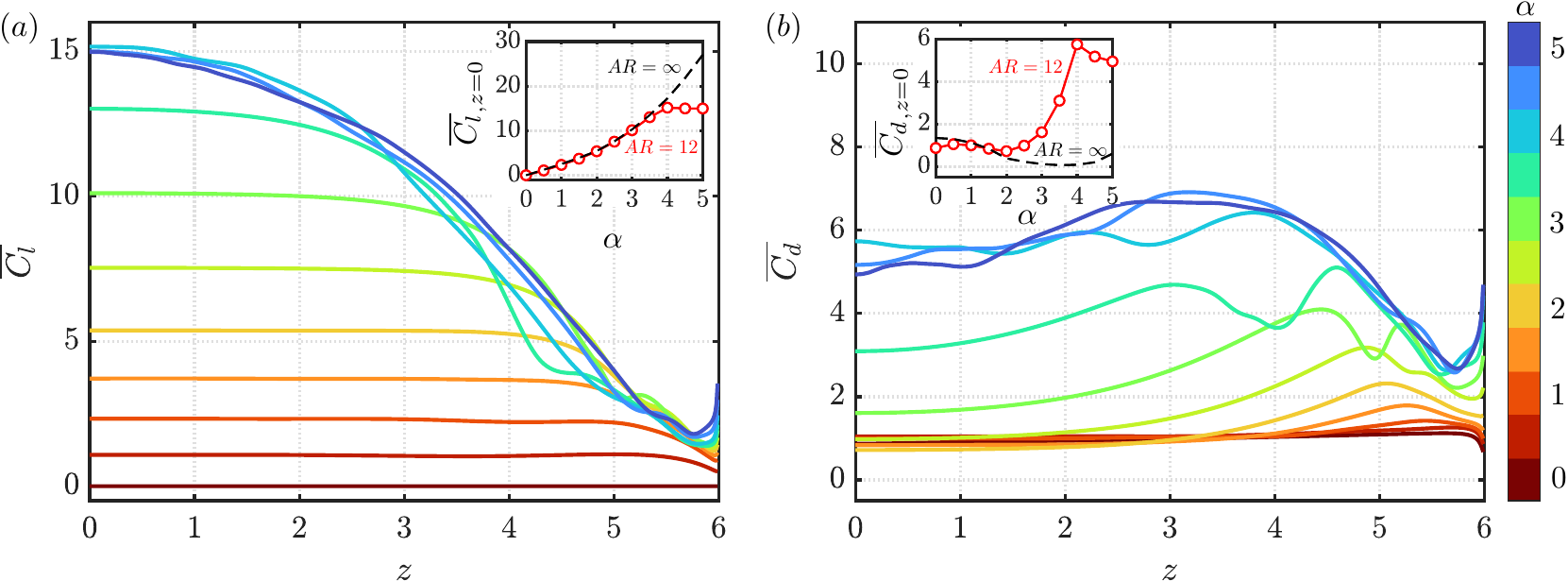}
\caption{Time-averaged sectional force coefficients for $AR=12$.  $(a)$ lift coefficients $\overline{C_l}$ and $(b)$ drag coefficients $\overline{C_d}$. Only half of the span is shown due to symmetry. In the inset plots, the sectional lift and drag coefficients for $AR=12$ at the midspan ($z=0$) are presented in red and compared to the 2-D results in black.}
\label{fig:SectionalForces}
\end{figure}

The time-averaged sectional lift and drag coefficients for $AR=12$ cylinders are presented in figure \ref{fig:SectionalForces}, providing quantitative insight into the three-dimensional free-end effects on force generation.
At low rotation rates, the sectional lift remains nearly uniform along the majority of the span, resembling two-dimensional flow conditions. 
However, a pronounced decline in $C_l$ occurs within the free-end region, where tip vortices and flow leakage reduces the Magnus-effect lift.
As $\alpha$ increases, two competing phenomena become apparent: first, the midspan sectional lift increases at a rate similar to that of two-dimensional flow; second, the zone affected by three-dimensional free-end effects expands inboard, progressively reducing the high-lift region.
By $\alpha \approx 4$, the free-end influence penetrates fully to inboard, causing even the middle sectional $\overline{C_l}_{,z=0}$ to fall below equivalent two-dimensional counterparts.
This observation explains the gradual saturation of the $\overline{C_L}$-$\alpha$ curves at high rotation rates, as observed in figure \ref{fig:ForceCoeffs}($a$).

Similarly, the sectional drag coefficients are strongly affected by free-end effects, as shown in figure \ref{fig:SectionalForces}($b$).
At low rotation rates, the drag remains relatively low and uniform along the span.
As the rotation rate $\alpha$ increases, regions near the free ends begin to exhibit elevated sectional drag, likely due to the development of intensified tip vortices and associated three-dimensional flow structures. 
This localized increase in drag progressively extends toward the midspan with further increases in $\alpha$.
Notably, the sectional drag at the midspan ($\overline{C_d}_{,z=0}$) begins to deviate significantly beyond $\alpha \approx 2$, indicating a growing influence of free-end effects even in regions that would otherwise approximate two-dimensional behavior.

\begin{figure}
\centering
\includegraphics[width=0.8\textwidth]{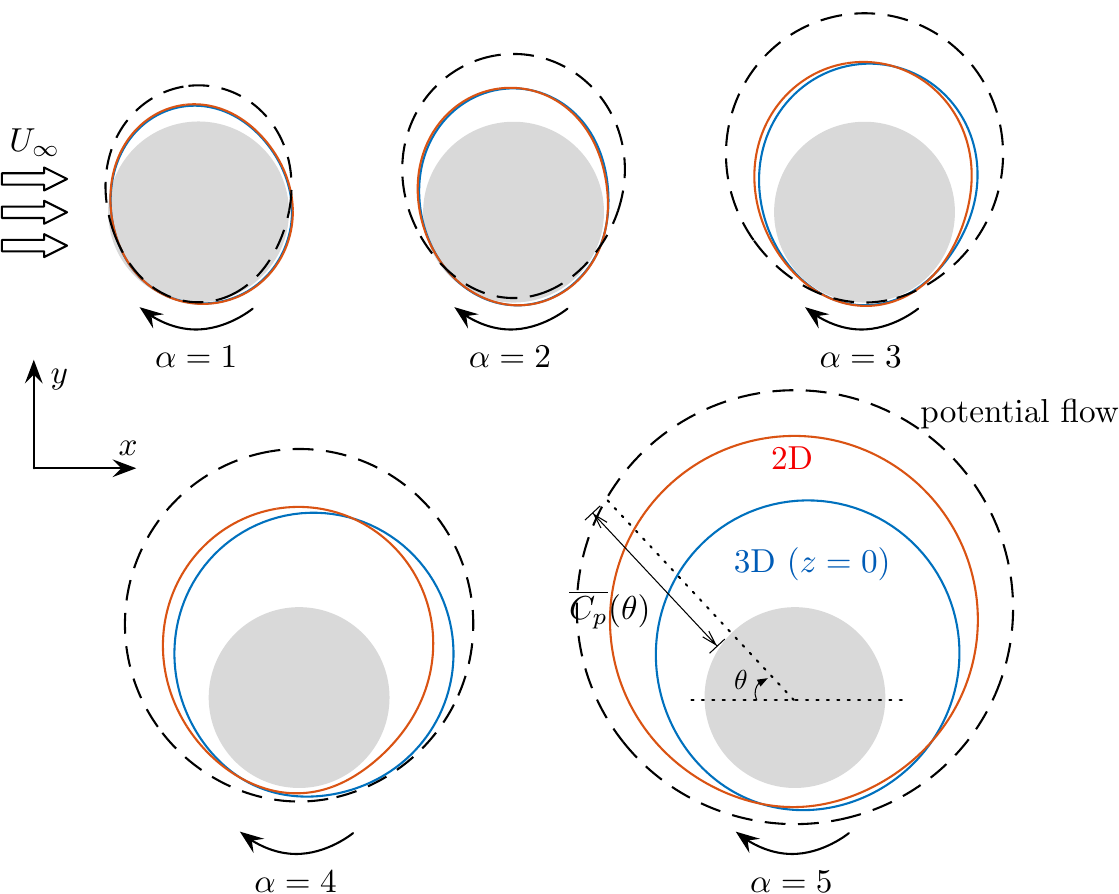}
\caption{Comparison of pressure distributions on rotating cylinder for $AR=12$. The gray circle represents the cylinder. Dashed lines: potential flow solution; orange lines: two-dimensional DNS results; blue lines: three-dimensional DNS results at mid-sections. The magnitudes of the pressure coefficients are shown as radial distances from the cylinder surface (negative pressure outside, positive pressure inside). }
\label{fig:AR12MidSectionPressureComparison}
\end{figure}

We further investigate the high sensitivity of the mid-section force coefficients to free-end effects by examining the pressure distributions over the cylinder surface at $AR=12$, as shown in figure \ref{fig:AR12MidSectionPressureComparison}.
According to potential flow theory, the pressure distribution around a rotating cylinder is given by:
\begin{equation}
C_p = 1-(2\sin\theta+\alpha)^2.
\label{equ:potentialCp}
\end{equation}
This pressure theoretical distribution is symmetric about the $y$-axis, resulting in zero net drag. 
Both the two-dimensional and three-dimensional simulations predict weaker suction on the retreating side compared to the potential flow solution, leading to a reduced lift coefficient. 
At low rotation rates ($\alpha = 1$--$2$), the $\overline{C_p}$ obtained the two- and three-dimensional simulations are nearly indistinguishable, indicating minimal three-dimensional influence at the midspan.
At intermediate rotation rates ($\alpha = 3$--$4$), the two-dimensional pressure distribution becomes increasingly symmetric about the $y$-axis. 
On the other hand, in the corresponding three-dimensional simulation, the pressure distribution skews rearward, introducing a non-negligible drag at the midspan.
At higher rotation rate ($\alpha = 5$), the pressure for the finite-$AR$ case becomes enclosed within that of the two-dimensional case, marking a clear departure from the idealized two-dimensional prediction.
These findings highlight the limitations of extrapolating two-dimensional results to finite-length rotating cylinders, particularly in high-$\alpha$ where free-end effects play a critical role.

\subsection{Effects of end plates}
\label{sec:endplates}

\begin{figure}
\centering
\includegraphics[width=0.8\textwidth]{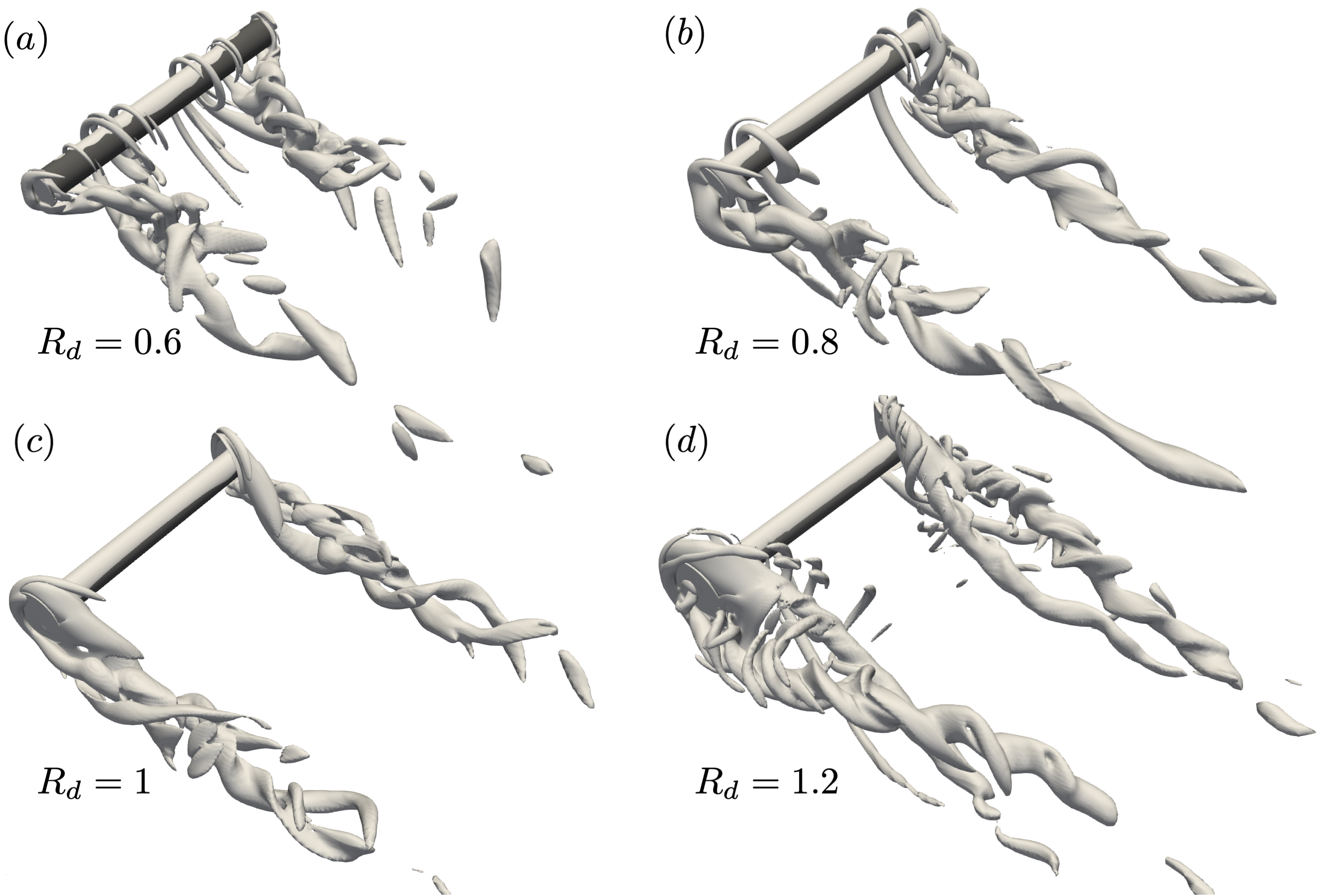}
\caption{Comparision of wake vortical structures ($Q=1$) with end plates at $(AR, \alpha)=(10,4)$. $(a)$ $R_{d}=0.6$, $(b)$ $R_{d}=0.8$, $(c)$ $R_{d}=1$ and $(d)$ $R_{d}=1.2$.}
\label{fig:EndPlateQCompare}
\end{figure}

We further examine the effectiveness of end plates on regulating the three-dimensional wake dynamics of finite rotating cylinders. 
This technique, also referred to as Thom disk \citep{Thom1934}, has been widely adopted in the Flettner rotors to enhance their aerodynamic efficiency \citep{thouault2012numerical,chen2023experimental}, but the wake dynamics has not been elucidated in detail.
In this study, we model the end plates as zero-thickness circular disks that rotate with the same speed as the cylinder.
The radius of the disk (normalized by the cylinder diameter) is varied from $R_{d}=0.6$ to 1.2.

Figure \ref{fig:EndPlateQCompare} shows a comparison of the wake vortical structures of rotating cylinders with different plate sizes for $(AR,\alpha)=(10,4)$.
Small end plates ($R_{d}=0.6$) minimally influence the three-dimensional wake flows, which remain characterized by strong tip vortices near the free end and Taylor-like vortices along the inboard span.
As the tip vortices are displaced further from the cylinder with increasing $R_d$, their structures become more complex. 
This is particularly evident for $R_d=1.2$, where two distinct vortex tubes form at the periphery of the rotating disk. 
The vortex tube on the retreating side tilts inboard, while its counterpart on the advancing side tilts outboard, forming a cross-finger configuration as they convect downstream.

\begin{figure}
\centering
\includegraphics[width=1\textwidth]{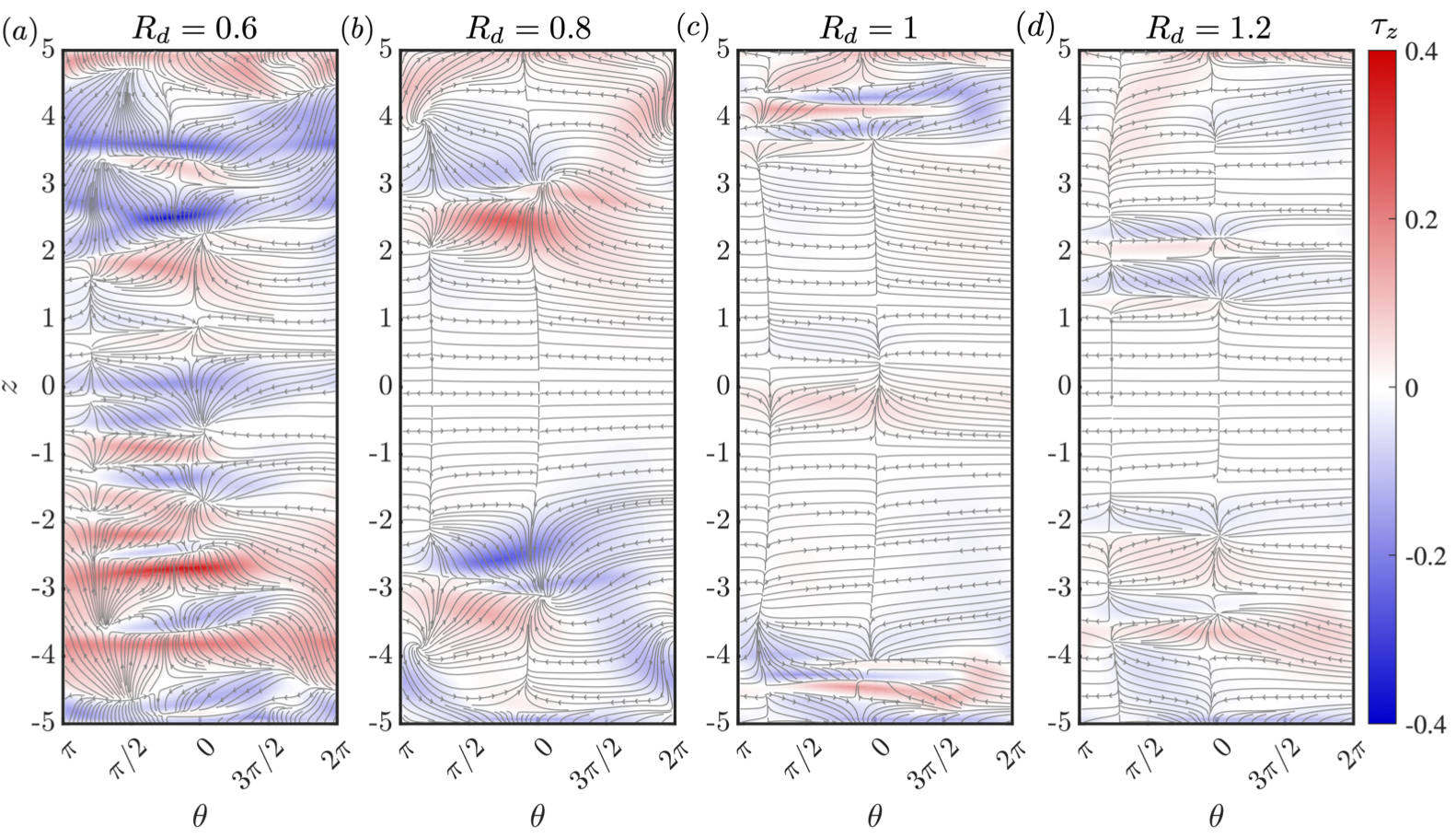}
\caption{Instantaneous skin friction line pattern for $(AR,\alpha)=(10,4)$. ($a$) $R_d=0.6$, ($b$) $R_d=0.8$, $(c)$ $R_d=1$ and $(d)$ $R_d=1.2$. The fields of $\tau_z$ is plotted to show the spanwise three dimensionality of the flow.}
\label{fig:EndPlateSFL}
\end{figure}

Despite the complex tip vortices, the incorporation of the end plates effectively suppresses the formation of the Taylor-like vortices, as evidenced in figure \ref{fig:EndPlateQCompare}($b,c,d$).
This phenomenon is more clearly illustrated in figure \ref{fig:EndPlateSFL}, which presents instantaneous skin friction line patterns for cases with $R_d$ ranging from 0.6 to 1.2. 
For $R_d=0.6$, the flow displays pronounced three-dimensionality due to the presence of Taylor-like vortices, manifesting as intricate patterns distributed across the entire cylinder surface.
As the end plate diameter increases, three-dimensional effects progressively shrink to the free ends, while the midspan region transitions toward quasi-two-dimensional behavior. 
Although residual spanwise flow persists near the midspan, the three-dimensionality becomes substantially less pronounced than that observed in the slip boundary case characterized by mode E, as demonstrated in figure \ref{fig:CompareWithSlipQ}($c$).

\begin{figure}
\centering
\includegraphics[width=1\textwidth]{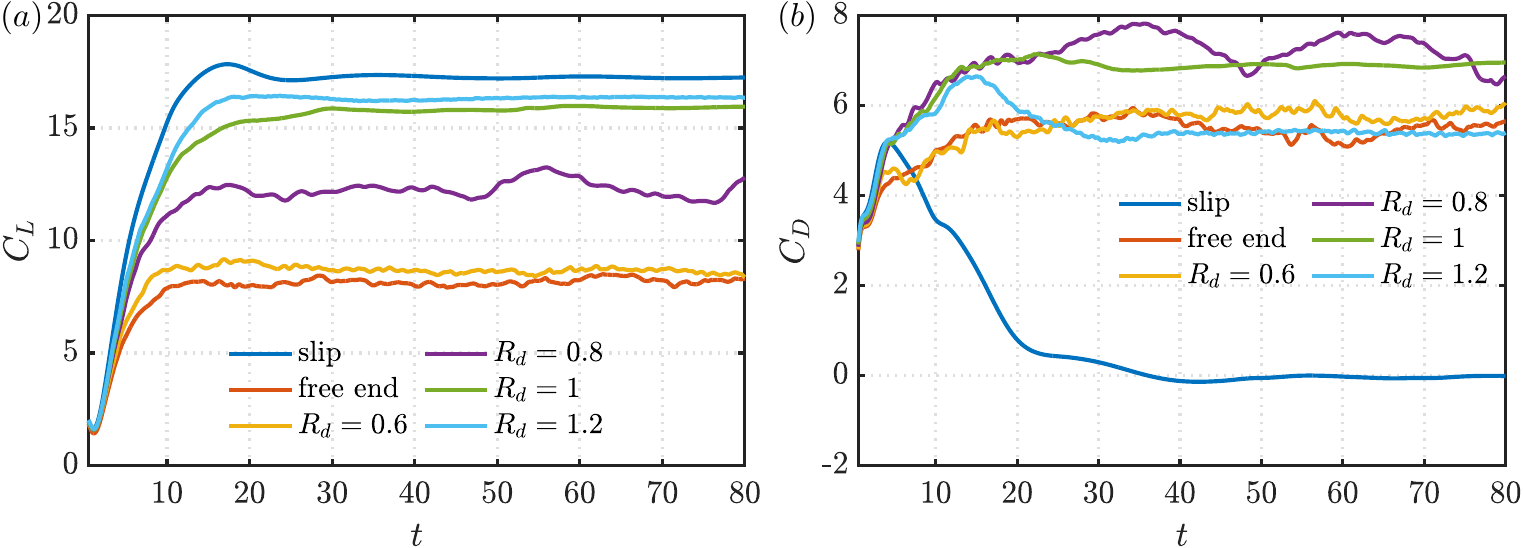}
\caption{Aerodynamic forces for flow over cylinders with end plates. $(a)$ Lift coefficient and ($b$) drag coefficient. The cases are with $(AR,\alpha)=(10,4)$.}
\label{fig:EndPlateCLCD}
\end{figure}

The aerodynamic performance of finite rotating cylinders can be substantially enhanced through the regulation of three-dimensionality by end plates, as demonstrated in figure \ref{fig:EndPlateCLCD}.
With increasing $R_d$, the lift coefficient exhibits monotonic growth, approaching values comparable to those observed in the two-dimensional cases. 
At $R_d=1.0$--$1.2$, the lift coefficient nearly doubles relative to the free end configuration.
The influence of end plates on drag coefficients is less pronounced than their effect on lift. 
Maximum drag occurs at $R_d=0.8$--$1.0$, while drag subsequently decreases to levels comparable to the free end case at $R_d=1.2$. 
Notably, even with the flow's two-dimensionalization achieved through end plates, drag coefficients remain considerably elevated compared to slip end conditions (for which $C_D\approx 0$).
Collectively, these findings demonstrate that end plates provide an effective means for regulating three-dimensional wake dynamics in rotating finite cylinders and enhancing aerodynamic performance.

\section{Conclusions}
\label{sec:conclusions}
We performed direct numerical simulations of three-dimensional flows over finite rotating circular cylinders at a Reynolds number of 150.
The aim is to characterize the effects of the free end on the three-dimensional wake dynamics of finite rotating cylinders.
An extensive parametric study was performed for aspect ratio $AR=2$--$12$ and rotation rate $\alpha=0$--$5$.

The most significant feature in the wakes of finite rotating cylinders is the formation of a pair of counter-rotating tip vortices. 
They impose downwash over the inboard region of the cylinder span, playing an important role in shaping the wake dynamics of finite rotating cylinders.
At low rotation rates, the wake exhibits periodic K\'arm\'an vortex shedding confined to the midspan, resembling typical bluff-body behavior.
For higher rotation rates, the weakened free shear layer and the intensified tip vortices both stabilize the wake.
Further increasing $\alpha$, the intensified tip vortices destabilize via mutual induction for the low-$AR$ cases, and self-induced instability for high-$AR$ cases.
At still higher $\alpha$, the Taylor-like vortices emerge as C-shaped structures from the free-end region, migrating toward the midspan driven by the self-induced velocity due to vortex-wall interaction.

The free-end effects significantly influence the aerodynamic forces of finite rotating cylinders. 
The lift coefficients increase with rotation rate but saturate at high $\alpha$ due to enhanced three-dimensional effects near the free ends.
The drag coefficients for finite-$AR$ cylinders are markedly higher than those from two-dimensional simulations. 
Pressure distributions at the midspan, despite being farthest from the free end, reveal persistent free-end effects, highlighting the limitations of two-dimensional simulations for assessing the aerodynamics of rotating cylinders in engineering applications.
Such free-end effects can be effectively suppressed by the end plates that position the tip vortices away from the cylinder.
With the two-dimensionalization of the wake along most of the span, the aerodynamic performance of the rotating cylinder is greatly improved. 

This study clarifies the role of free-end effects in shaping three-dimensional wake dynamics behind finite rotating cylinders.
The insights obtained from this study are critical for the understanding of the more complex wake flows at higher Reynolds number, as some of the coherent vortices at high $Re$ share similar core structures with the analogous vortical wakes of low $Re$ flow. 
This is especially true for the tip vortices that remain coherent across a wide range of Reynolds numbers.
These fundamental findings provide a robust foundation for advancing the design and optimization of rotating cylinder systems across diverse engineering applications, from wind-assisted propulsion to flow control technologies.

\section*{Acknowledgment}
The financial supports from the National Key R\&D Program of China (No. 2023YFE0120000), the Guangdong Basic and Applied Basic Research Foundation (No. 2023A1515240054), the Program for Intergovernmental International S\&T Cooperation Projects of Shanghai Municipality, China (No. 24510711100), and National Natural Science Foundation of China (Nos. 12202271 and 52478535) are acknowledged.

 \bibliographystyle{unsrtnat}
 \bibliography{refs}

\begin{thebibliography}{67}
\providecommand{\natexlab}[1]{#1}
\providecommand{\url}[1]{\texttt{#1}}
\expandafter\ifx\csname urlstyle\endcsname\relax
  \providecommand{\doi}[1]{doi: #1}\else
  \providecommand{\doi}{doi: \begingroup \urlstyle{rm}\Url}\fi

\bibitem[Magnus(1853)]{Magnus1853}
G.~Magnus.
\newblock Ueber die abweichung der geschosse, und: Ueber eine auffallende
  erscheinung bei rotirenden körpern.
\newblock \emph{Annalen der Physik}, 164\penalty0 (1):\penalty0 1--29, 1853.

\bibitem[Borg(1986)]{Borg1986Magnus}
J.~F. Borg.
\newblock Magnus effect: An overview of its past and future practical
  applications. volume i.
\newblock Technical Report AD-A165 902, Naval Sea Systems Command, Washington,
  DC, 1986.

\bibitem[Anderson(2011)]{anderson2011fundamentals}
J.~Anderson.
\newblock \emph{Fundamentals of Aerodynamics}.
\newblock McGraw Hill, 6 edition, 2011.

\bibitem[Flettner(1925)]{flettner1925}
A.~Flettner.
\newblock The {F}lettner rotor ship.
\newblock \emph{Eng.}, 19:\penalty0 117--120, 1925.

\bibitem[Seifert(2012)]{seifert2012review}
J.~Seifert.
\newblock A review of the magnus effect in aeronautics.
\newblock \emph{Prog. Aerosp. Sci.}, 55:\penalty0 17--45, 2012.

\bibitem[Sedaghat(2014)]{sedaghat2014magnus}
A.~Sedaghat.
\newblock Magnus type wind turbines: Prospectus and challenges in design and
  modelling.
\newblock \emph{Renew. Energy}, 62:\penalty0 619--628, 2014.

\bibitem[Schulmeister et~al.(2017)Schulmeister, Dahl, Weymouth, and
  Triantafyllou]{schulmeister2017flow}
J.~C. Schulmeister, J.~M. Dahl, G.~D. Weymouth, and M.~S. Triantafyllou.
\newblock Flow control with rotating cylinders.
\newblock \emph{J. Fluid Mech.}, 825:\penalty0 743--763, 2017.

\bibitem[Fan et~al.(2020)Fan, Yang, Wang, Triantafyllou, and
  Karniadakis]{fan2020reinforcement}
D.~Fan, L.~Yang, Z.~Wang, M.~S. Triantafyllou, and G.~E. Karniadakis.
\newblock Reinforcement learning for bluff body active flow control in
  experiments and simulations.
\newblock \emph{Proc. Natl. Acad. Sci.}, 117\penalty0 (42):\penalty0
  26091--26098, 2020.

\bibitem[Yu et~al.(2020)Yu, Ping, Liu, Zhu, Wang, Bao, Zhou, Han, and
  Xu]{yu2020turbulent}
Z.~Yu, H.~Ping, X.~Liu, H.~Zhu, R.~Wang, Y.~Bao, D.~Zhou, Z.~Han, and H.~Xu.
\newblock Turbulent wake suppression of circular cylinder flow by two small
  counter-rotating rods.
\newblock \emph{Phys. Fluids}, 32\penalty0 (11), 2020.

\bibitem[Bao et~al.(2022)Bao, Ping, Zhu, Zhou, Yang, and Han]{bao2022laminar}
Y.~Bao, H.~Ping, H.~Zhu, D.~Zhou, Y.~Yang, and Z.~Han.
\newblock Laminar wake suppression of airfoil by rotating rod at low {R}eynolds
  number.
\newblock \emph{Phys. Rev. Fluids}, 7\penalty0 (3):\penalty0 034102, 2022.

\bibitem[Wu et~al.(2022)Wu, R{\"o}mer, Axtmann, and Rist]{wu2022transition}
Y.~Wu, T.~R{\"o}mer, G.~Axtmann, and U.~Rist.
\newblock Transition mechanisms in a boundary layer controlled by rotating
  wall-normal cylindrical roughness elements.
\newblock \emph{J. Fluid Mech.}, 945:\penalty0 A20, 2022.

\bibitem[Stojkovi\'{c} et~al.(2002)Stojkovi\'{c}, Breuer, and
  Durst]{Stojkovic2002pof}
D.~Stojkovi\'{c}, M.~Breuer, and F.~Durst.
\newblock Effect of high rotation rates on the laminar flow around a circular
  cylinder.
\newblock \emph{Phys. Fluids}, 14\penalty0 (9):\penalty0 3160--3178, 09 2002.

\bibitem[Stojkovi\'{c} et~al.(2003)Stojkovi\'{c}, Sch{\"o}n, Breuer, and
  Durst]{Stojkovic2003pof}
D.~Stojkovi\'{c}, P.~Sch{\"o}n, M.~Breuer, and F.~Durst.
\newblock On the new vortex shedding mode past a rotating circular cylinder.
\newblock \emph{Phys. Fluids}, 15\penalty0 (5):\penalty0 1257--1260, 05 2003.

\bibitem[Mittal and Kumar(2003)]{mittal2003flow}
S.~Mittal and B.~Kumar.
\newblock Flow past a rotating cylinder.
\newblock \emph{J. Fluid Mech.}, 476:\penalty0 303--334, 2003.

\bibitem[Padrino and Joseph(2006)]{PADRINO_JOSEPH_2006}
J.~C. Padrino and D.~D. Joseph.
\newblock Numerical study of the steady-state uniform flow past a rotating
  cylinder.
\newblock \emph{J. Fluid Mech.}, 557:\penalty0 191–223, 2006.

\bibitem[Pralits et~al.(2010)Pralits, Brandt, and
  Giannetti]{pralits2010instability}
J.~O. Pralits, L.~Brandt, and F.~Giannetti.
\newblock Instability and sensitivity of the flow around a rotating circular
  cylinder.
\newblock \emph{J. Fluid Mech.}, 650:\penalty0 513--536, 2010.

\bibitem[Rao et~al.(2013{\natexlab{a}})Rao, Leontini, Thompson, and
  Hourigan]{Rao2013jfm_a}
A.~Rao, J.~S. Leontini, M.~C. Thompson, and K.~Hourigan.
\newblock Three-dimensionality in the wake of a rapidly rotating cylinder in
  uniform flow.
\newblock \emph{J. Fluid Mech.}, 730:\penalty0 379–391, 2013{\natexlab{a}}.

\bibitem[Thompson et~al.(2014)Thompson, Rao, Leontini, and
  Hourigan]{Thompson_et_al_2014}
M.~C. Thompson, A.~Rao, J.~S. Leontini, and K.~Hourigan.
\newblock The existence of multiple solutions for rotating cylinder flows.
\newblock In \emph{19th Australasian Fluid Mechanics Conference}, Melbourne,
  Australia, December 8-11 2014. RMIT University.

\bibitem[Sierra et~al.(2020)Sierra, Fabre, Citro, and
  Giannetti]{sierra2020bifurcation}
J.~Sierra, D.~Fabre, V.~Citro, and F.~Giannetti.
\newblock Bifurcation scenario in the two-dimensional laminar flow past a
  rotating cylinder.
\newblock \emph{J. Fluid Mech.}, 905:\penalty0 A2, 2020.

\bibitem[Br{\o}ns(2021)]{brons2021organizing}
M.~Br{\o}ns.
\newblock The organizing centre for the flow around rapidly spinning cylinders.
\newblock \emph{J. Fluid Mech.}, 906:\penalty0 F1, 2021.

\bibitem[Radi et~al.(2013)Radi, Thompson, Rao, Hourigan, and
  Sheridan]{Radi2013JFM}
A.~Radi, M.~C. Thompson, A.~Rao, K.~Hourigan, and J.~Sheridan.
\newblock Experimental evidence of new three-dimensional modes in the wake of a
  rotating cylinder.
\newblock \emph{J. Fluid Mech.}, 734:\penalty0 567–594, 2013.

\bibitem[Rao et~al.(2013{\natexlab{b}})Rao, Leontini, Thompson, and
  Hourigan]{Rao2013jfm_b}
A.~Rao, J.~Leontini, M.~C. Thompson, and K.~Hourigan.
\newblock Three-dimensionality in the wake of a rotating cylinder in a uniform
  flow.
\newblock \emph{J. Fluid Mech.}, 717:\penalty0 1–29, 2013{\natexlab{b}}.

\bibitem[Pralits et~al.(2013)Pralits, Giannetti, and Brandt]{pralits2013three}
J.~O. Pralits, F.~Giannetti, and L.~Brandt.
\newblock Three-dimensional instability of the flow around a rotating circular
  cylinder.
\newblock \emph{J. Fluid Mech.}, 730:\penalty0 5--18, 2013.

\bibitem[Rao et~al.(2015)Rao, Radi, Leontini, Thompson, Sheridan, and
  Hourigan]{rao2015review}
A.~Rao, A.~Radi, J.~S. Leontini, M.~C. Thompson, J.~Sheridan, and K.~Hourigan.
\newblock A review of rotating cylinder wake transitions.
\newblock \emph{J. Fluids Struct.}, 53:\penalty0 2--14, 2015.

\bibitem[Navrose et~al.(2015)Navrose, Meena, and Mittal]{navrose2015three}
Navrose, J.~Meena, and S.~Mittal.
\newblock Three-dimensional flow past a rotating cylinder.
\newblock \emph{J. Fluid Mech.}, 766:\penalty0 28--53, 2015.

\bibitem[Williamson(1996)]{williamson1996ARFM}
C.~H.~K. Williamson.
\newblock Vortex dynamics in the cylinder wake.
\newblock \emph{Annu. Rev. Fluid Mech.}, 28:\penalty0 477--539, 1996.
\newblock ISSN 1545-4479.

\bibitem[Taylor(1923)]{Taylor1923PTRS}
G.~I. Taylor.
\newblock Stability of a viscous liquid contained between two rotating
  cylinders.
\newblock \emph{Phil. Trans. R. Soc. Lond. Ser. A}, 102\penalty0
  (718):\penalty0 541--542, 1923.

\bibitem[Drazin and Reid(2004)]{Drazin_Reid_2004}
P.~G. Drazin and W.~H. Reid.
\newblock \emph{Hydrodynamic Stability}.
\newblock Cambridge University Press, 2 edition, 2004.

\bibitem[Slaouti and Gerrard(1981)]{slaouti1981experimental}
A.~Slaouti and J.~H. Gerrard.
\newblock An experimental investigation of the end effects on the wake of a
  circular cylinder towed through water at low {R}eynolds numbers.
\newblock \emph{J. Fluid Mech.}, 112:\penalty0 297--314, 1981.

\bibitem[Ramberg(1983)]{ramberg1983effects}
S.~E. Ramberg.
\newblock The effects of yaw and finite length upon the vortex wakes of
  stationary and vibrating circular cylinders.
\newblock \emph{J. Fluid Mech.}, 128:\penalty0 81--107, 1983.

\bibitem[Williamson(1989)]{williamson1989oblique}
C.~H.~K. Williamson.
\newblock Oblique and parallel modes of vortex shedding in the wake of a
  circular cylinder at low reynolds numbers.
\newblock \emph{J. Fluid Mech.}, 206:\penalty0 579--627, 1989.

\bibitem[Park and Lee(2000)]{park2000free}
C.-W. Park and S.-J. Lee.
\newblock Free end effects on the near wake flow structure behind a finite
  circular cylinder.
\newblock \emph{J. Wind Eng. Ind. Aerod.}, 88\penalty0 (2-3):\penalty0
  231--246, 2000.

\bibitem[Roh and Park(2003)]{roh2003vortical}
S.~Roh and S.~Park.
\newblock Vortical flow over the free end surface of a finite circular cylinder
  mounted on a flat plate.
\newblock \emph{Exp. Fluids}, 34\penalty0 (1):\penalty0 63--67, 2003.

\bibitem[Inoue and Sakuragi(2008)]{inoue2008vortex}
O.~Inoue and A.~Sakuragi.
\newblock Vortex shedding from a circular cylinder of finite length at low
  {R}eynolds numbers.
\newblock \emph{Phys. Fluids}, 20\penalty0 (3), 2008.

\bibitem[Wang and Zhou(2009)]{WANG_ZHOU_2009}
H.~F. Wang and Y.~Zhou.
\newblock The finite-length square cylinder near wake.
\newblock \emph{J. Fluid Mech.}, 638:\penalty0 453–490, 2009.

\bibitem[Krajnovi\'c(2011)]{KRAJNOVIC_2011}
S.~Krajnovi\'c.
\newblock Flow around a tall finite cylinder explored by large eddy simulation.
\newblock \emph{J. Fluid Mech.}, 676:\penalty0 294–317, 2011.

\bibitem[Sumner(2013)]{sumner2013flow}
D.~Sumner.
\newblock Flow above the free end of a surface-mounted finite-height circular
  cylinder: {A} review.
\newblock \emph{J. Fluids Struct.}, 43:\penalty0 41--63, 2013.

\bibitem[Mittal et~al.(2021)Mittal, Pandi, and Hore]{mittal2021cellular}
S.~Mittal, J.~S.~S. Pandi, and M.~Hore.
\newblock Cellular vortex shedding from a cylinder at low {R}eynolds number.
\newblock \emph{J. Fluid Mech.}, 915, 2021.

\bibitem[Cao et~al.(2022)Cao, Tamura, Zhou, Bao, and Han]{cao2022jfm}
Y.~Cao, T.~Tamura, D.~Zhou, Y.~Bao, and Z.~Han.
\newblock Topological description of near-wall flows around a surface-mounted
  square cylinder at high {R}eynolds numbers.
\newblock \emph{J. Fluid Mech.}, 933:\penalty0 A39, 2022.

\bibitem[Zhang et~al.(2023)Zhang, Bao, Han, and Zhou]{zhang2023OE}
K.~Zhang, Y.~Bao, Z.~Han, and D.~Zhou.
\newblock End boundary effects on wakes dynamics of inclined circular
  cylinders.
\newblock \emph{Ocean Eng.}, 269:\penalty0 113543, 2023.

\bibitem[Taira and Colonius(2009)]{taira2009three}
K.~Taira and T.~Colonius.
\newblock Three-dimensional flows around low-aspect-ratio flat-plate wings at
  low {R}eynolds numbers.
\newblock \emph{J. Fluid Mech.}, 623:\penalty0 187--207, 2009.

\bibitem[DeVoria and Mohseni(2017)]{DeVoria_Mohseni_2017}
A.~C. DeVoria and K.~Mohseni.
\newblock On the mechanism of high-incidence lift generation for steadily
  translating low-aspect-ratio wings.
\newblock \emph{J. Fluid Mech.}, 813:\penalty0 110–126, 2017.

\bibitem[Zhang et~al.(2020{\natexlab{a}})Zhang, Hayostek, Amitay, Burtsev,
  Theofilis, and Taira]{zhang2020laminar}
K.~Zhang, S.~Hayostek, M.~Amitay, A.~Burtsev, V.~Theofilis, and K.~Taira.
\newblock Laminar separated flows over finite-aspect-ratio swept wings.
\newblock \emph{J. Fluid Mech.}, 905:\penalty0 R1, 2020{\natexlab{a}}.

\bibitem[Zhang et~al.(2020{\natexlab{b}})Zhang, Hayostek, Amitay, He,
  Theofilis, and Taira]{zhang2020formation}
K.~Zhang, S.~Hayostek, M.~Amitay, W.~He, V.~Theofilis, and K.~Taira.
\newblock On the formation of three-dimensional separated flows over wings
  under tip effects.
\newblock \emph{J. Fluid Mech.}, 895:\penalty0 A9, 2020{\natexlab{b}}.

\bibitem[Zhang and Taira(2022)]{zhang2022laminar}
K.~Zhang and K.~Taira.
\newblock Laminar vortex dynamics around forward-swept wings.
\newblock \emph{Phys. Rev. Fluids}, 7\penalty0 (2):\penalty0 024704, 2022.

\bibitem[Pandi and Mittal(2023)]{pandi2023JFM}
J.~S.~S. Pandi and S.~Mittal.
\newblock Streamwise vortices, cellular shedding and force coefficients on
  finite wing at low {R}eynolds number.
\newblock \emph{J. Fluid Mech.}, 958:\penalty0 A10, 2023.

\bibitem[Smith and Taira(2024)]{smith2024effect}
L.~Smith and K.~Taira.
\newblock The effect of {R}eynolds number on the separated flow over a
  low-aspect-ratio wing.
\newblock \emph{J. Fluid Mech.}, 992:\penalty0 R2, 2024.

\bibitem[Zhu et~al.(2024)Zhu, Wang, and Liu]{Zhu_Wang_Liu_2024}
Y.~Zhu, J.~Wang, and J.~Liu.
\newblock Tip effects on three-dimensional flow structures over
  low-aspect-ratio plates: mechanisms of spanwise fluid transport.
\newblock \emph{J. Fluid Mech.}, 983:\penalty0 A35, 2024.

\bibitem[Mittal(2004)]{mittal2004JAM}
S.~Mittal.
\newblock Three-dimensional instabilities in flow past a rotating cylinder.
\newblock \emph{J. Appl. Mech.}, 71:\penalty0 89--95, 2004.

\bibitem[Yang et~al.(2023)Yang, Wang, Guo, and Zhang]{yang2023JFM}
Y.~Yang, C.~Wang, R.~Guo, and M.~Zhang.
\newblock Numerical analyses of the flow past a short rotating cylinder.
\newblock \emph{J. Fluid Mech.}, 975:\penalty0 A15, 2023.

\bibitem[Massaro et~al.(2024)Massaro, Karp, Jansson, Markidis, and
  Schlatter]{massaro2024direct}
D.~Massaro, M.~Karp, N.~Jansson, S.~Markidis, and P.~Schlatter.
\newblock Direct numerical simulation of the turbulent flow around a {F}lettner
  rotor.
\newblock \emph{Sci. Rep.}, 14:\penalty0 3004, 2024.

\bibitem[Bordogna et~al.(2019)Bordogna, Muggiasca, Giappino, Belloli, Keuning,
  Huijsmans, and Van't~Veer]{bordogna2019experiments}
G.~Bordogna, S.~Muggiasca, S.~Giappino, M.~Belloli, J.~A. Keuning, R.~H.~M.
  Huijsmans, and A.~P. Van't~Veer.
\newblock Experiments on a {F}lettner rotor at critical and supercritical
  {R}eynolds numbers.
\newblock \emph{J. Wind Eng. Ind. Aerodyn.}, 188:\penalty0 19--29, 2019.

\bibitem[Ma et~al.(2022)Ma, Liu, Jia, Jin, and Ma]{ma2022jweia}
W.~Ma, J.~Liu, Y.~Jia, L.~Jin, and X.~Ma.
\newblock The aerodynamic forces and wake flow of a rotating circular cylinder
  under various flow regimes.
\newblock \emph{J. Wind Eng. Ind. Aerod.}, 224:\penalty0 104977, 2022.
\newblock ISSN 0167-6105.

\bibitem[Bordogna et~al.(2020)Bordogna, Muggiasca, Giappino, Belloli, Keuning,
  and Huijsmans]{BORDOGNA2020JWEIA}
G.~Bordogna, S.~Muggiasca, S.~Giappino, M.~Belloli, J.A. Keuning, and R.H.M.
  Huijsmans.
\newblock The effects of the aerodynamic interaction on the performance of two
  {F}lettner rotors.
\newblock \emph{J. Wind Eng. Ind. Aerodyn.}, 196:\penalty0 104024, 2020.

\bibitem[Chen et~al.(2023)Chen, Wang, and Liu]{chen2023experimental}
W.~Chen, H.~Wang, and X.~Liu.
\newblock Experimental investigation of the aerodynamic performance of flettner
  rotors for marine applications.
\newblock \emph{Ocean Eng.}, 281:\penalty0 115006, 2023.

\bibitem[Schlichting and Truckenbrodt(2013)]{schlichting2013aerodynamik}
H.~Schlichting and E.~A. Truckenbrodt.
\newblock \emph{Aerodynamik des Flugzeuges: Erster Band Grundlagen aus der
  Strömungsmechanik Aerodynamik des Tragflügels (Teil I)}.
\newblock Springer-Verlag, 2013.

\bibitem[Lighthill(1963)]{Lighthill1963}
M.~J. Lighthill.
\newblock Attachment and separation in three-dimensional flows. laminar
  boundary layers theory.
\newblock \emph{Sect. 927 II2.6, ed. L Rosenhead, New York: Oxford Univ.
  Press}, pages 72--82, 1963.

\bibitem[Tobak and Peake(1982)]{Tobak1982}
M.~Tobak and D.~J. Peake.
\newblock Topology of three-dimensional separated flows.
\newblock \emph{Annu. Rev. Fluid Mech.}, 14\penalty0 (1):\penalty0 61--85,
  1982.

\bibitem[Délery(2001)]{Delery2001}
J.~M. Délery.
\newblock Robert legendre and henri werlé: Toward the elucidation of
  three-dimensional separation.
\newblock \emph{Annual Review of Fluid Mechanics}, 33\penalty0 (33):\penalty0
  129--154, 2001.

\bibitem[Davey(1961)]{Davey1961}
A.~Davey.
\newblock Boundary-layer flow at a saddle point of attachment.
\newblock \emph{J. Fluid Mech.}, 10\penalty0 (4):\penalty0 593--610, 1961.

\bibitem[Taneda(1952)]{taneda1952studies}
S.~Taneda.
\newblock Studies on wake vortices ({I}). {A}n experimental study on the
  structure of the vortex street behind a circular cylinder of finite length.
\newblock \emph{Res. Inst. Appl. Mech}, 1:\penalty0 131--143, 1952.

\bibitem[Levold(2012)]{levold2012viscous}
P.~Levold.
\newblock Viscous flow around finite length circular cylinder.
\newblock Master's thesis, Norwegian University of Science and Technology,
  2012.

\bibitem[Schmid(2010)]{SCHMID_2010}
P.~J. Schmid.
\newblock Dynamic mode decomposition of numerical and experimental data.
\newblock \emph{J. Fluid Mech.}, 656:\penalty0 5–28, 2010.

\bibitem[Milne-Thomson(1968)]{milnethomson1968theoretical}
L.~M. Milne-Thomson.
\newblock \emph{Theoretical Hydrodynamics}.
\newblock Dover, London, 5th edition, 1968.

\bibitem[Prandtl(1925)]{prandtl1925magnuseffekt}
L.~Prandtl.
\newblock Magnuseffekt und windkraftschiff.
\newblock \emph{Sci. Nat.}, 13\penalty0 (6):\penalty0 93--108, 1925.

\bibitem[Thom(1934)]{Thom1934}
A.~Thom.
\newblock Effects of discs on the air forces on a rotating cylinder.
\newblock Tech. Rep. 1623, Aeronautical Research Committee, 1934.

\bibitem[Thouault et~al.(2012)Thouault, Breitsamter, Adams, Seifert,
  Badalamenti, and Prince]{thouault2012numerical}
N.~Thouault, C.~Breitsamter, N.~A Adams, J.~Seifert, C.~Badalamenti, and S.~A.
  Prince.
\newblock Numerical analysis of a rotating cylinder with spanwise disks.
\newblock \emph{AIAA J.}, 50\penalty0 (2):\penalty0 271--283, 2012.

\end{thebibliography}

\end{document}